\documentclass[11pt,onecolumn,draft]{IEEEtran} 
\usepackage{cite}
\usepackage{psfig}
\usepackage{subfigure}
\usepackage{amsmath}
\usepackage{bm}
\usepackage{amssymb}

\newcommand{\bb}[1]{{\mathbf{#1}}}

\newcommand{\nn}{\nonumber}

\DeclareMathOperator{\diag}{diag} \DeclareMathOperator{\tr}{tr}

\newtheorem{theorem}{Theorem}

\newtheorem{lemma}{Lemma}
\newtheorem{definition}{Definition}
\begin{document}

\title{Asymptotic Eigenvalue Moments of Wishart-Type Random Matrix Without Ergodicity in One Channel Realization}

\author{Chien-Hwa Hwang\\
Institute of Communications Engineering\\
\& Department of Electrical Engineering,\\
National Tsing Hua University,\\
Hsinchu, Taiwan.\\
E-mail: chhwang@ee.nthu.edu.tw}

\maketitle

\begin{abstract}
Consider a random matrix whose variance profile is random. This random matrix is ergodic in one channel realization if, for each column and row, the empirical distribution of the squared magnitudes of elements therein converges to a nonrandom distribution. In this paper, noncrossing partition theory is employed to derive expressions for several asymptotic eigenvalue moments (AEM) related quantities of a large Wishart-type random matrix $\bb H\bb H^\dag$ when $\bb H$ has a random variance profile and is nonergodic in one channel realization. It is known the empirical eigenvalue moments of $\bb H\bb H^\dag$ are dependent (or independent) on realizations of the variance profile of $\bb H$ when $\bb H$ is nonergodic (or ergodic) in one channel realization. For nonergodic $\bb H$, the AEM can be obtained by i) deriving the expression of AEM in terms of the variance profile of $\bb H$, and then ii) averaging the derived quantity over the ensemble of variance profiles. Since the AEM are independent of the variance profile if $\bb H$ is ergodic, the expression obtained in i) can also serve as the AEM formula for ergodic $\bb H$ when any realization of variance profile is available. 
\end{abstract}

\begin{keywords}
Random matrix, Wishart matrix, variance profile, asymptotic eigenvalue moments (AEM), noncrossing partition.
\end{keywords}

\section{Introduction}

Consider the linear vector memoryless model $\bb y=\bb H\bb
x+\bb w$, where $\bb x$, $\bb y$ and $\bb w$ are the input vector,
output vector and additive white Gaussian noise (AWGN),
respectively, and $\bb H$ denotes the random channel matrix
independent of $\bb w$. Entries of the matrix $\bb H$ depend on the actual application, and the linear model is characterized by the joint distribution of entries in $\bb H$.
It is known that, if the elements of a sized $N\times K$ random matrix $\bb H$ are independent and identically distributed (i.i.d.) zero-mean random variables having a common variance $1/N$,
then the empirical distribution of the eigenvalues of $\bb H\bb H^\dag$
converges almost surely (a.s.) to the Mar\u{c}enko-Pastur law \cite{marcenko67} when
$K,N\rightarrow\infty$ with a finite ratio $K/N$.
However, in many applications, entries of $\bb H$ have unequal variances, i.e. independent but non-identically distributed (i.n.d.). Examples include direct sequence-code division multiple access (DS-CDMA) and multicarrier (MC)-CDMA systems with frequency-flat/selective fading, multiaccess system with antenna diversity,
and so forth. Moreover, in a multiple-input-multiple-output (MIMO) system with spatially correlated fading, the channel matrix with correlated
elements can be transformed to another matrix having the same
asymptotic eigenvalue distribution (AED) but with independent and
unequal-variances components \cite{sayeed02,tulino05_1,weichselberger06}. In these cases, explicit expressions for the AED of $\bb H\bb H^\dag$ rarely exist. Fortunately, in many occasions, the AED can be characterized in a  way that several performance metrics such as capacity, minimum mean-square-error (MMSE), multiuser efficiency, optimum output signal-to-interference-plus-noise ratio (SINR), etc., can be evaluated through solving nonlinear equations numerically. A number of works follow this line include \cite{girko90,shlyakhtenko96,tse99,kiran00,shamai01,li04,tulino05,tulino05_1}, and a more complete list can be found in \cite{tulino2004}. They all start from the celebrated Girko's theorem [\citenum{girko90}, Corollary 10.1.2], where the Stieltjes transform for the AED of $\bb H\bb H^\dag$ is obtained as an integral of the solution to a fixed-point equation depending on the variances of entries in $\bb H$.

In the literature, the asymptotic eigenvalue moments (AEM) of a random matrix are much less explored than AED. This is mainly because a moment based method reveals little underlying behavior and the combinatorial arguments involved are frequently horrible. Nevertheless, it appears unlikely to dispense with the method since, for example, there has been no alternative way of proving the behavior of the extreme eigenvalues \cite{silverstein99}. Some application cases of AEM are exemplified below.
\begin{itemize}
\item Let the moments $m_k=\textrm{E}\{\lambda^k\},k=1,2,\cdots$ be available for $K,N\to\infty$ and finite $K/N$, where $\lambda$ is the random variable governing the eigenvalues of $\bb H\bb H^\dag$. These moments are called the AEM of $\bb H\bb H^\dag$. When the Carleman's criterion $\sum_{k=1}^\infty m_{2k}^{-1/(2k)}=\infty$ holds, the moment sequence $\{m_k\}$ uniquely determines a distribution (AED of $\bb H\bb H^\dag$) \cite{carleman22}; thus, one can use $\{m_k\}$ to characterize the system. Suppose that the goal is to evaluate the expectation of a certain function $g$ of the random variable $\lambda$, e.g., $g(\lambda)=\log_2 (1+\gamma\lambda)$ for the ergodic capacity and $g(\lambda)=1/(1+\gamma\lambda)$ for MMSE under the input signal-to-noise ratio (SNR) $\gamma$. The Gauss quadrature rule method \cite{golub69} can be adopted to achieve the goal, where the expectation $\textrm{E}\{g(\lambda)\}$ is expressed as a linear combination of samples of $g(\lambda)$, and $\{m_k\}$ is used to determine the coefficients in the combination and the points to be sampled.

\item The AEM is relevant to the design and analysis of a reduced-rank MMSE receiver \cite{moshavi96,muller2001,li04,tulino01,li01,cottatellucci02,hachem02}, which is able to reach the performance of a linear MMSE receiver with a lower computational cost.

\item The AEM plays an important role in free probability theory \cite{voiculescu92}, which has been lately used as a powerful tool to analyze complicated wireless communication models \cite{tse99_1,evans00,biglieri02,muller02,debbah03,muller04,peacock06,ryan07}.

\end{itemize}

Let $H_{i,j}$ denote the $(i,j)$-th entry of matrix $\bb H$. The variance profile of a random matrix $\bb H$ is a matrix whose $(i,j)$-th entry is the variance of $H_{i,j}$. Consider the case that the variance profile of $\bb H$ is random as well. Then $\bb H$ is said to be ergodic in one channel realization if, for each column and row, the empirical distribution of $|H_{i,j}|^2$ therein converges to a nonrandom distribution [\citenum{tulino05}, Defintion 2.3]. It is known, if $\bb H$ is ergodic in one channel realization, the empirical eigenvalue distribution of $\bb H\bb H^\dag$ converges a.s. to a nonrandom limit whose Stieltjes transform can be given by [\citenum{girko90}, Corollary 10.1.2], where it can be seen the AED as well as AEM of $\bb H\bb H^\dag$ are independent of any specific realization of variance profile of $\bb H$. However, in many practical applications, $\bb H$ is nonergodic in one channel realization. In this paper, we assume that, although $\bb H$ is \textit{nonergodic in one channel realization}, it is \textit{ergodic in the time domain}. Specifically, we suppose that the variance profile of $\bb H$ is controlled by some random variables (denoted by $\pmb\theta$); conditioned on a certain variance profile, the conditional AEM converges asymptotically to quantities that are functions of realizations of $\pmb\theta$. Since $\bb H$ is ergodic in the time domain, further averaging over the ensemble of the random vector $\pmb\theta$ yields the unconditional AEM.\footnote{In some occasions, instead of performing the averaging process on AEM, it is more reasonable that we compute the average of the quantity of interest, e.g. spectral efficiency, coefficients of a reduced-rank MMSE receiver, etc. However, these quantities are generally related to $\pmb\theta$ in such a complicated manner that it is intractable to evaluate the average. Thus, performing expectation over the AEM and then using it to obtain the interested quantity appears to be a feasible alternative.}

Most previous works employing asymptotic results of eigenvalues of $\bb H\bb H^\dag$ either have the results conditioned on a certain realization of $\bb H$'s variance profile (deterministic variance profile) or consider random variance profile but assume $\bb H$ is ergodic in one channel realization. Examples of nonergodic $\bb H$ are given below.
A one-shot asynchronous CDMA system is considered in \cite{kiran00}, where the
empirical distribution of $|H_{i,j}|^2$ in a column is a function of the random relative delay and received power of the user corresponding to that column. To find the optimal output SINR, the statistics of the random variables controlling the variance profile matrix are incorporated into the fixed-point equation governing the Stieltjes transform. In \cite{tulino05}, a number of performance measures of an MC-CDMA system are analyzed, where a situation is taken into account that the frequency-selective fading channel is nonergodic in the frequency domain but ergodic in the time domain\footnote{In this case, the randomness in the spectrum of the channel matrix due to the realization of the fading process does not vanish asymptotically. Thus, the performance measures converge asymptotically to quantities that are functions of the fading realization, and further averaging over the fading ensemble yields the ergodic asymptotic performance measures when the fading is temporally ergodic.}. The performance measures conditioned on a specific variance profile matrix are averaged over the ensemble of the fading process to yield the ergodic performance measures. In \cite{sayeed02,weichselberger06}, mappings between the scattering radio environment of MIMO and a spare virtual channel matrix (or called coupling matrix) are established. The spare channel matrix contains many zero elements, where the number of nonzero entries indicates the channel degree of freedom, and the distribution of non-zero elements in the grid corresponds to the radio environments.
It is desirable to construct a statistic model for the virtual channel matrix to govern the ensemble radio channels over time and space.

Some related works about AEM computation with i.n.d. entries of $\bb H$ are summarized, and short comments are made to compare them with our work. Expressions for AEM of $\bb H\bb H^\dag$ are derived in \cite{li04} for the cases that entries of $\bb H$ are zero-mean and i.n.d. and $\bb H$ is ergodic in one channel realization. The expressions therein are recursion based. They are obtained by expanding the fixed-point equation of the Stieltjes transform of AED as a power series. As a Stieltjes transform can be formulated as power series with moments as the coefficients, AEM are identified by equating the two power series expressions (one by expanding the fixed-point equation of the Stieltjes transform, and the other by a power series with moments as the coefficients) and comparing the coefficients of every degree at either side of the equality. In the current paper, for a particular realization of $\bb H$'s variance profile, we derive conditional AEM expressions in both direct and recursive forms. For the latter, our expression exhibits a simpler form than that in \cite{li04}. Recursive forms have the advantage of lower computation complexity when the moment order is high; however, it is inconvenient to perform averaging process on them to obtain unconditional AEM formulas.  In \cite{shlyakhtenko96}, a fundamental observation is made that limits of Gaussian band matrices are operator-valued semicircular elements; thus operator-valued free probability can be used for determining their eigenvalue distribution.
The works of \cite{far06,far08} employ the result of \cite{shlyakhtenko96} and take an operator-valued free probability approach to calculate the limits of the eigenvalue distributions
of a number of block matrices, including Wishart-type block matrices. In the intermediate process, the operator-valued moments in a recurrence relation can be obtained. By using a trick of putting $\bb H$ and $\bb H^\dag$ in off-diagonal blocks of a $2\times 2$ block matrix $\bb X$ (diagonal blocks are zero matrices), a Gaussian matrix $\bb X$ is obtained whose AEM (conditional) has a simple mapping with the AEM of $\bb H\bb H^\dag$. However, the AEM obtained in this way is also in recurrence, and it is inconvenient to extend to unconditional AEM.

The rest of this paper is organized as follows. Section II lays down some necessary definitions. In Section III, expressions of AEM in both direct and recursive forms are derived given a certain realization of the variance profile of $\bb H$. In Section IV, AEM formulas are obtained by means of the direct-form result in Section III, and two statistic models of nonergodic $\bb H$ are investigated.
Applications are given in Section V, and this paper is concluded in Section VI.

\section{Preliminaries}\label{section:prelim}

\begin{definition}[Empirical Distribution]
The empirical distribution of the vector $[v_1,v_2,\cdots,v_N]$ is given as
$$
F(x)=\frac{1}{N}\sum_{i=1}^N\textrm{u}(x-v_i),
$$
where $\textrm{u}(\cdot)$ is the unit step function.\hfill{\small $\blacksquare$}
\end{definition}

Consider an $N\times K$ matrix $\bb H=\bb M\odot\bb S$, where $\odot$ denotes the element-wise matrix product, and $\bb M$ and $\bb S$ are independent $N\times K$ random matrices. The entries
of $\bb S$ are arbitrarily distributed zero-mean i.i.d. complex random variables with variance $1/N$. We denote the $(i,j)$-th entries of $\bb H$ and $\bb M$ by $H_{i,j}$ and $M_{i,j}$, respectively.
\begin{definition}[Ergodicity in One Channel Realization\cite{tulino05}]\label{def:channelProfile}
Let $\lfloor\cdot\rfloor$ denote the closest smaller integer. For a given $x\in[0,1)$, let the empirical distribution of
$$
[|M_{\lfloor xN\rfloor,1}|^2,|M_{\lfloor xN\rfloor,2}|^2,\cdots,|M_{\lfloor xN\rfloor,K}|^2]
$$
converge to $F_x(\cdot)$ when $K,N\to\infty$ and $K/N\to\beta$ a finite constant; for a given $y\in[0,1)$, let the empirical distribution of
$$
[|M_{1,\lfloor yK\rfloor}|^2,|M_{2,\lfloor yK\rfloor}|^2,\cdots,|M_{N,\lfloor yK\rfloor}|^2]
$$
converge to $F_y(\cdot)$ as $K,N\to\infty$ and $K/N\to\beta$.
If the asymptotic empirical distributions $F_x$ and $F_y$, $x,y\in[0,1)$,
are a.s. nonrandom limits, then the random matrix $\bb H$ is called ergodic in one channel realization. \hfill{\small $\blacksquare$}
\end{definition}

Conditioned on a realization of $\bb M$, the variance of $H_{i,j}$ is equal to $|M_{i,j}|^2/N$. The variance profile of $\bb H$ given a certain realization of $\bb M$ is defined below.
\begin{definition}[Conditional Variance Profile]\label{def:vp}
Consider a realization of random matrix $\bb M$. Let
$$
V(i,j):=|M_{i,j}|^2.
$$
For each $N$ and $K$ with their ratio $K/N=\beta$, let $v^{(N)}:[0,1)\times[0,1)\to\mathbb{R}^+$ be a function given by
\begin{equation}\label{eq:mouse0918}
v^{(N)}(x,y)=V(i,j),\quad\frac{i-1}{N}\leq x<\frac{i}{N},\quad \frac{j-1}{K}\leq y<\frac{j}{K}.
\end{equation}
Assume that the sequence $\{v^{(N)}(x,y)\}_{N=1}^\infty$ converges to a limiting bounded function $v(x,y)$. Then, $v(x,y)$ is referred to as the conditional variance profile of $\bb H$ given $\bb M$. Furthermore, for a given $1\leq k\leq K$, let $v_k^{(N)}$ be a function defined in $[0,1)$ such that
\begin{equation}\label{eq:mouse0922}
v_k^{(N)}(x)=V(i,k),\quad\frac{i-1}{N}\leq x<\frac{i}{N}.
\end{equation}
The limit of $v_k^{(N)}$, denoted as $v_k(x)$, is called the conditional variance profile of the $k$-th column of $\bb H$ given $\bb M$. \hfill{\small $\blacksquare$}
\end{definition}

By the Girko's theorem [\citenum{girko90}, Corollary 10.1.2], conditioned on $\bb M$, the empirical eigenvalue distribution of $\bb H\bb H^\dag$ converges a.s. to a nonrandom limit whose Stieltjes transform ${\cal S}(z)$ can be expressed as
\begin{equation}\label{eq:mouse0926}
{\cal S}(z)=\int_0^1 u(x,z)\textrm{d}x,
\end{equation}
where $u(x,z)$ satisfies the equation
\begin{equation}\label{eq:tiger0926}
u(x,z)=\dfrac{1}{1+z\beta\int_0^1\frac{v(x,y)}{1+z\int_0^1 u(w,z)v(w,y)\textrm{d}w}\textrm{d}y}.
\end{equation}
It is readily seen from (\ref{eq:mouse0926}) and (\ref{eq:tiger0926}) that, when $\bb H$ is ergodic in one channel realization, the conditional AED (as well as the quantities of our interest, AEM) of $\bb H\bb H^\dag$ given $\bb M$ is invariant to the conditional variance profile $v(x,y)$ and hence the realization of $\bb M$.
However, in many practical applications, $\bb H$ is nonergodic in one channel realization, and the AEM of $\bb H\bb H^\dag$ depend on the specific realization of $\bb M$. This is the case to be investigated in this paper, where we assume that, although $\bb H$ is \textit{nonergodic in one channel realization}, it is \textit{ergodic in the time domain}. Specifically, we suppose that the realization of $\bb M$ is controlled by a set of random variables in $\pmb\theta$; conditioned on a certain $\bb M$, the AEM converge asymptotically to quantities that are functions of the realization of $\pmb\theta$. We assume that $\bb H$ is ergodic in the time domain; thus, further averaging over the ensemble of $\pmb\theta$ yields the unconditional AEM.

The statistics of random matrix $\bb M$ are explicitly given below. We represent a random process by a function $Z(u,t)$ of two variables, where $u$ is a point that varies over the sample space $\cal U$, and $t$ is a point over an index set $\cal T$.
In the current context, we suppose that a particular column $k$ of $\bb M$ has a different statistical property from all of the others. This occurs when, for example, user $k$ is the desired user in the processing of the linear model $\bb y=\bb H\bb x+\bb w$.
\begin{definition}\label{def:4}
Consider a realization of random matrix $\bb M$ and let $V(i,j)=|M_{i,j}|^2$. For each given $j\in[1,K]\setminus\{k\}$, $\{V(i,j):1\leq i\leq N\}$ is a sample function of the discrete-time stochastic process $Z_j(u,t)$ whose moments up to a certain order exist, and $\{Z_j(u,t):j\in[1,K]\setminus\{k\}\}$ is a collection of i.i.d. random processes. Furthermore, for the $k$-th column of $\bb M$, $\{V(i,k):1\leq i\leq N\}$ is a realization of the random process $Z_k(u,t)$.\footnote{The relation between $Z_j(u,t)$, $j\neq k$, and $Z_k(u,t)$ may be stated below. Let the former and the latter be defined by a set of random variables $\pmb\theta$ and $\pmb{\theta}_k$, respectively. In many occasions, $\pmb{\theta}_k\subset\pmb\theta$, and $Z_k(u,t)$ is obtained by fixing the random variables of $\pmb\theta\setminus\pmb{\theta}_k$ by known deterministic values. The rationale of this lays on that the $k$-th column corresponds to the desired user and thus more information is available.}\hfill{\small $\blacksquare$}
\end{definition}

Due to i.i.d. of $Z_j(u,t)$'s, we will use statistics of $Z(u,t)$ to represent those of $Z_j(u,t)$'s.
By (\ref{eq:mouse0918}), for any $y\in\{(j-1)/K:1\leq j\leq K,j\neq k\}$, we can obtain a continuous-time random process, in which $\{v^{(N)}(x,y):0\leq x<1\}$ is its realization. This continuous-time random process inherits properties from its discrete-time counterpart. That is, its moments up to a certain order exist, processes corresponding to distinct index $y$ are i.i.d., and the statistics of these i.i.d. processes are characterized by the process $z^{(N)}(u,t)$. When $N\to\infty$, the limit of $z^{(N)}(u,t)$ is represented by $z(u,t)$.
A continuous-time counterpart $z_k^{(N)}(u,t)$ of $Z_k(u,t)$ can be found using (\ref{eq:mouse0922}), and the limit of $z_k^{(N)}(u,t)$ is symbolized by $z_k(u,t)$ when $N\to\infty$.

We define three random sequences related to moments of the Wishart-type random matrix $\bb H\bb H^\dag$:
\begin{eqnarray}
\mu_m^{(N)}&:=&N^{-1}\tr\{(\bb H\bb
H^\dag)^m\},\label{eq:mouse0427}\\
\eta_{m,k}^{(N)}&:=&((\bb H^\dag\bb
H)^m)_{k,k},\label{eq:cow0427}\\
\delta_{m,k}^{(N)}&:=&\bb h_k^\dag(\bb H_{\sim k}\bb H^\dag_{\sim
k})^{m}\bb h_k\label{eq:tiger0427},
\end{eqnarray}
where the superscript $^{(N)}$ denotes that the quantity is
evaluated when $\bb H$ has $N$ number of rows, $\bb h_k$ is the
$k$-th column of $\bb H$, and $\bb H_{\sim k}$ is $\bb H$ with $\bb
h_k$ removed. Suppose that $\mu_m^{(N)}$, $\eta_{m,k}^{(N)}$
and $\delta_{m,k}^{(N)}$ converge to $\mu_m$, $\eta_{m,k}$ and
$\delta_{m,k}$, respectively, when $K,N\rightarrow\infty$ and $K/N\rightarrow\beta$. The importance of
these asymptotic moment-related quantities are explained below.
\begin{itemize}
\item The quantity $\mu_m$ is the general definition for the $m$-th AEM of
$\bb H\bb H^\dag$.

\item Given the model $\bb y=\bb H\bb x+\bb w$ described in
Introduction, assume the $k$-th entry of $\bb x$ is of interest, and
$\bb H$ is known to the receiver.
Then $\eta_{m,k}$ is the quantity concerned with the detection of the desired symbol.

\item The quantity $\delta_{m,k}$ is often used
in the design and performance evaluation of a reduced-rank receiver
with the received signal model $\bb y=\bb H\bb x+\bb w$
\cite{honig2001,tulino01,muller2001,li01,li04}. In \cite{moshavi96,cottatellucci05_1}, $\eta_{m,k}$ is used for the same goal.
\end{itemize}

In the following, some definitions regarding set partition theory
are summarized. The properties that will be used in our later
derivations are also provided.
\begin{definition}[Noncrossing
Partition\cite{kreweras72,speicher06}]\label{def:3} Let $S$ be a finite totally
ordered set.
\begin{enumerate}
\item We call $\varpi=\{B_1,\cdots,B_j\}$ a partition of the set $S$ if
and only if $B_i$, $1\leq i\leq j$, are pairwise disjoint, non-empty
subsets of $S$ such that $B_1\cup\cdots\cup B_j=S$. We call
$B_1,\cdots,B_j$ the blocks of $\varpi$, and $|\varpi|$ and $|B_i|$
represent the number of blocks in $\varpi$ and the number of
elements in $B_i$, respectively. The blocks $B_1,\cdots,B_j$ are
ordered according to the minimum element in each block. That is, the
minimum element in $B_k$ is smaller than that in $B_l$ if $k<l$. For
$p\in S$, ${\cal B}_{\varpi}(p)$ denotes the index of the block that
$p$ belongs to under the partition $\varpi$. For example, if $p\in
B_i$ under $\varpi$, then ${\cal B}_{\varpi}(p)=i$.

\item The collection of all partitions of $S$ can be viewed as a
partially ordered set (poset) in which the partitions are ordered by
\textit{refinement}: if $\varpi,\sigma$ are two partitions of $S$,
we have $\varpi\leq \sigma$ if each block of $\varpi$ is contained
in a block of $\sigma$. For example, when $S=\{1,2,3,4,5,6,7\}$, we
have
$\{\{1\},\{2,5\},\{3,4\},\{6\},\{7\}\}\leq\{\{1,3,4\},\{2,5\},\{6,7\}\}$.

\item A partition $\varpi$ of the set $S$ is called crossing if there
exist $p_1<q_1<p_2<q_2$ in $S$ such that $p_1$ and $p_2$ belong to
one block, and $q_1$ and $q_2$ belong to another.
If $\varpi$ is not crossing, then it is called noncrossing.
\hfill{\small $\blacksquare$}
\end{enumerate}
\end{definition}

The set of all noncrossing partitions of $S$ is denoted by $NC(S)$.
In the special case of $S=\{1,\cdots,m\}$, we denote this by
$NC(m)$.

\begin{definition}[Kreweras Complementation Map\cite{kreweras72,speicher06}]
Consider numbers $\overline{1},\overline{2},\cdots,\overline{m}$, and we form a
totally ordered set
$\{1,\overline{1},2,\overline{2},\cdots,m,\overline{m}\}$
by interlacing them with $1,2,\cdots,m$. Let $\varpi\in
NC(\{1,\cdots,m\})$. Then its Kreweras complementation map
$K(\varpi):NC(\{1,\cdots,m\})\rightarrow
NC(\{\overline{1},\cdots,\overline{m}\})$ is defined to be the
largest element among those $\sigma\in
NC(\{\overline{1},\cdots,\overline{m}\})$ such that
$\varpi\cup\sigma$ belongs to $NC
(\{1,\overline{1},\cdots,m,\overline{m}\})$, where "largest" is in
the sense described in item 2) of Definition~\ref{def:3}.
$\hfill{\small \blacksquare}$
\end{definition}

In this paper, noncrossing partition and Kreweras complementation
map are employed to derive the limits of sequences
(\ref{eq:mouse0427})--(\ref{eq:tiger0427}). The same tool has been employed by \cite{li01,xiao00,xiao05} in obtaining AEM when entries of $\bb H$ are i.i.d. A convenient
representation of a noncrossing partition $\varpi$ as well as its
Kreweras complementation map $K(\varpi)$ is the $K$-graph detailed in Appendix~\ref{appendix:K-graph}.

\section{Asymptotic Moments Conditioned on a Certain Realization}

Consider an $N\times K$ random matrix $\bb H$ as described in
Section~\ref{section:prelim}. We define
\begin{eqnarray}\label{eq:sheep0822}
&&\hat{\mu}_m:=\mathop{\lim_{K,N\to\infty}}_{K/N\to\beta} \textrm{E}\{\mu_m^{(N)}|v^{(N)}(x,y)\},\quad
\hat{\eta}_{m,k}:=\mathop{\lim_{K,N\to\infty}}_{K/N\to\beta}\textrm{E}\{\eta_{m,k}^{(N)}|v^{(N)}(x,y)\},\nn\\
&\mbox{and}& \quad
\hat{\delta}_{m,k}:=\mathop{\lim_{K,N\to\infty}}_{K/N\to\beta}\textrm{E}\{\delta_{m,k}^{(N)}|v^{(N)}(x,y)\}.
\end{eqnarray}
In this section, we compute the above three quantities that are conditioned on a specific $v^{(N)}(x,y)$, and we show that $\hat{\mu}_m$, $\hat{\eta}_{m,k}$, and $\hat{\delta}_{m,k}$ are the limiting values of the sequences given in (\ref{eq:mouse0427}), (\ref{eq:cow0427}) and (\ref{eq:tiger0427}), respectively, for each particular realization of $v^{(N)}(x,y)$.

\begin{theorem}\label{theorem:1}
Conditioned on $v^{(N)}(x,y)$, the random sequences $\mu_m^{(N)}$, $\eta_{m,k}^{(N)}$ and $\delta_{m,k}^{(N)}$
converge a.s. to $\hat{\mu}_m$, $\hat{\eta}_{m,k}$ and $\hat{\delta}_{m,k}$, respectively, as $K,N\to\infty$ and $K/N\to\beta$. We have
\begin{equation}
\hat{\mu}_m=\mathop{\sum_{\varpi\in
NC(m)}}_{\varpi=\{B_1,\cdots,B_{|\varpi|}\}}\beta^{|\varpi|}\textrm{E}\left\{
\prod_{i=1}^{|\varpi|}\prod_{s_i\in \{{\cal
B}_{K(\varpi)}(t):t\in B_i\} }v(X_{s_i},Y_i)\right\}\label{eq:mouse1114},
\end{equation}
\begin{eqnarray}\label{eq:mouse0501}
\hat{\eta}_{m,k}=\mathop{\sum_{\varpi\in
NC(m)}}_{\varpi=\{B_1,\cdots,B_{|\varpi|}\}} \beta^{|\varpi|-1}\textrm{E}\left\{\prod_{s_1\in \{{\cal
B}_{K(\varpi)}(t):t\in B_1\}
}v_k(X_{s_1})\prod_{i=2}^{|\varpi|}\prod_{s_i\in \{{\cal B}_{K(\varpi)}(t):t\in B_i\}
}v(X_{s_i},Y_i)\right\},
\end{eqnarray}
and
\begin{equation}\label{eq:tiger0827}
\hat{\delta}_{m,k}=\mathop{\sum_{\varpi\in
NC(m)}}_{\varpi=\{B_1,\cdots,B_{|\varpi|}\}}
\beta^{|\varpi|}\textrm{E}
\left\{v_k(X_1)\prod_{i=1}^{|\varpi|}
\prod_{s_i\in \{{\cal B}_{K(\varpi)}(t):t\in B_i\}
}v(X_{s_i},Y_i)\right\},
\end{equation}
where the expectations are with respect to continuous i.i.d. random variables $X_1,\cdots,X_{m-|\varpi|+1},$ $Y_1,\cdots,Y_{|\varpi|}$ uniformly distributed in $[0,1)$.
\end{theorem}
\begin{proof}
The proof is provided in Appendix~\ref{appendix:proof1}, which relies on the representation of $K$-graph.
\end{proof}

Note that, since the AEM of $\bb H\bb H^\dag$ are independent of realizations of $\bb M$ if $\bb H$ is ergodic in one channel realization, the derived conditional AEM in Theorem~\ref{theorem:1} can also serve as the AEM formula for ergodic $\bb H$ when any realization of $\bb M$ is available.
Issues regarding the complexity reduction in obtaining $\hat{\mu}_m$, $\hat{\eta}_{m,k}$ and $\hat{\delta}_{m,k}$ above are addressed before we move on to taking ensemble average of conditional AEM according to the distribution of $v^{(N)}(x,y)$. To evaluate the expressions in (\ref{eq:mouse1114})--(\ref{eq:tiger0827}), it requires $(m+1)$-dimensional integrals with the $m$-th Catalan number of times\footnote{The number of noncrossing partitions of an $m$-element set is equal to the $m$-th Catalan number.}.
When the order $m$ of the moment is high, the computation cost is huge. If the conditional variance profile $v(x,y)$ possesses a special structure, the AEM expressions can be simplified through properties of noncrossing partitions. For instance, if $v(x,y)$ is decomposable as $g(x)h(y)$, where $g(x)$ and $h(y)$ are nonnegative valued functions with span $[0,1)$, then formulas given in Theorem~\ref{theorem:1} admit simpler forms. The complexity can also be reduced by resorting to recursions in conditional AEM expressions. We explore these two ways of complexity reduction in Theorems~\ref{corollary:1} and \ref{theorem:recursion}.

The simpler form of $\hat{\mu}_m$ when $v(x,y)=g(x)h(y)$ has been obtained in Theorem~3 of \cite{li04}. For completeness, formulas of
$\hat{\mu}_m$, $\hat{\eta}_{m,k}$ and $\hat{\delta}_{m,k}$ given that $v(x,y)=g(x)h(y)$ are all provided in the following theorem.
\begin{theorem}\label{corollary:1}
If the conditional variance profile is decomposable as $v(x,y)=g(x)h(y)$ and $v_k(x)=\alpha_k g(x)$, then
\begin{enumerate}
\item $\hat{\mu}_m$ can be given by
\begin{eqnarray}
&&\sum_{l=1}^m
\beta^{l}\mathop{\sum_{b_1+\cdots+b_l=m}}_{b_1\geq\cdots\geq
b_l\geq
1}\mathop{\sum_{c_1+\cdots+c_{m-l+1}=m}}_{c_1\geq
\cdots\geq c_{m-l+1}\geq 1}
\dfrac{m(m-l)!(l-1)!}{f(b_1,\cdots,b_l)
f(c_1,\cdots,c_{m-l+1})}\nn\\
&&\hspace{1cm}\times\prod_{i=1}^{l}
\textrm{E}\{h(X)^{b_i}\}\prod_{j=1}^{m-l+1}
\textrm{E}\{g(X)^{c_j}\} ,\label{eq:tiger0828}
\end{eqnarray}
where
$f(n_1,\cdots,n_i)=\prod_{k\geq 1} p_k!$ with $p_k$ the number of elements in $(n_1,\cdots,n_i)$ that are equal to $k$, and $X$ is a continuous uniform random variable in $[0,1)$.

\item $\hat{\eta}_{m,k}$ can be given by
\begin{eqnarray}
&&\sum_{l=1}^m
\beta^{l-1}\mathop{\sum_{b_1+\cdots+b_l=m}}_{b_1\geq\cdots\geq
b_l\geq
1}\mathop{\sum_{c_1+\cdots+c_{m-l+1}=m}}_{c_1\geq
\cdots\geq c_{m-l+1}\geq 1}
\dfrac{(m-l)!(l-1)!}{f(b_1,\cdots,b_l)
f(c_1,\cdots,c_{m-l+1})}\nn\\
&&\hspace{1cm}\times \prod_{i=1}^l
\textrm{E}\{ h(X)^{b_i}\} \prod_{j=1}^{m-l+1}\textrm{E}\{
g(X)^{c_j}\}
\sum_{n=1}^l \frac{b_n\alpha_k^{b_n}}{\textrm{E}\{h(X)^{b_n}\}}.\label{eq:mouse0829}
\end{eqnarray}

\item $\hat{\delta}_{m,k}$ can be given by
\begin{eqnarray}
&&\alpha_k\sum_{l=1}^{m}
\beta^{l}\mathop{\sum_{b_1+\cdots+b_l=m}}_{b_1\geq\cdots\geq
b_l\geq
1}\mathop{\sum_{c_1+\cdots+c_{m-l+1}=m}}_{c_1\geq
\cdots\geq c_{m-l+1}\geq 1}
\dfrac{(m-l)!(l-1)!}{f(b_1,\cdots,b_l)
f(c_1,\cdots,c_{m-l+1})}\nn\\
&&\hspace{.5cm}\times\prod_{i=1}^{l}\textrm{E}\{
h(X)^{b_i}\} \prod_{j=1}^{m-l+1}\textrm{E}\{
g(X)^{c_j}\}\sum_{n=1}^{m-l+1}\frac{c_n \textrm{E}\{g(X)^{c_n+1}\}}{\textrm{E}\{ g(X)^{c_n}\}}.\label{eq:cow0829}
\end{eqnarray}

\end{enumerate}

\end{theorem}

\begin{proof}
See Appendix~\ref{appendix:3}.
\end{proof}

In the following theorem, recursive formulas of $\hat{\mu}_m$, $\hat{\eta}_{m,k}$ and $\hat{\delta}_{m,k}$ are provided. They have simpler forms compared with those given in  \cite{li04} for $\hat{\mu}_m$ and $\hat{\delta}_{m,k}$ (Theorems~1 and 2 of \cite{li04} with notations $\mu_m$ and $\delta_{m,k}$) by means of expanding the Stieltjes transform equations (\ref{eq:mouse0926}) and (\ref{eq:tiger0926}). For some particular variance profile, it is possible to obtain closed-form conditional AEM expressions by symbolic operations of scientific computation softwares such as MATLAB.
\begin{theorem}\label{theorem:recursion}
The a.s. limiting value of $\textrm{E}\{\mu_m^{(N)}|v^{(N)}(x,y)\}$ can be given as
\begin{equation}\label{eq:horse0819}
\hat{\mu}_m=\beta\cdot\textrm{E}\{\tilde{\mu}_m(X,Y)\},
\end{equation}
where the expectation is over i.i.d. random variables $X$ and $Y$ that are uniformly
distributed in $[0,1)$, and $\tilde{\mu}_m(x,y)$ can be obtained from the
recursion
\begin{equation}\label{eq:snake0819}
\tilde{\mu}_m(x,y)=v(x,y)\left(\beta\sum_{i=1}^{m-1}\textrm{E}\{\tilde{\mu}_{i-1}(X,y)\}\textrm{E}\{\tilde{\mu}_{m-i}(x,Y)\}+
\textrm{E}\{\tilde{\mu}_{m-1}(X,y)\}\right),\quad m\geq 1
\end{equation}
with $\tilde{\mu}_0(x,y)=1$.
Also, we have
\begin{equation}\label{eq:horse0822}
\hat{\eta}_{m,k}=\textrm{E}\{\tilde{\eta}_{m,k}(X)\}\quad\mbox{and}\quad \hat{\delta}_{m,k}=\beta\cdot \textrm{E}\{v_k(X)\tilde{\mu}_m(X,Y)\},
\end{equation}
where $X$ and $Y$ are i.i.d. uniform random variables in $[0,1)$, and $\tilde{\eta}_{m,k}(x)$ can be obtained recursively from
\begin{equation}\label{eq:tiger0822}
\tilde{\eta}_{m,k}(x)=v_k(x)\left(\beta\sum_{i=1}^{m-1}\hat{\eta}_{i-1,k}\textrm{E}\{\tilde{\mu}_{m-i}(x,Y)\}+
\hat{\eta}_{m-1,k}\right),\quad m\geq 1
\end{equation}
with $\tilde{\eta}_{0,k}(x)=1$.
\end{theorem}
\begin{proof}
See Appendix~\ref{appendix:recursion}.
\end{proof}

\section{Unconditional AEM}\label{section:4}

Unconditional AEM can be obtained as the ensemble averages of conditional AEM $\hat{\mu}_m$, $\hat{\eta}_{m,k}$ and $\hat{\delta}_{m,k}$ according to the distribution of $v(x,y)$. It is readily seen that performing the above operation to recursion based AEM expressions is inconvenient. Thus, it is suggested that results in Theorems~\ref{theorem:1} and \ref{corollary:1} are employed. We denote the ensemble averages of $\hat{\mu}_m$, $\hat{\eta}_{m,k}$ and $\hat{\delta}_{m,k}$ by $\mu_m$, $\eta_{m,k}$, and $\delta_{m,k}$, respectively.

Recall, in Definition~\ref{def:4}, we describe that, for any $j\in[1,K]\setminus\{k\}$, $\{V(i,j):1\leq i\leq N\}$ is a realization of random process $Z_j(u,t)$, and $Z_j(u,t)$'s are i.i.d. for distinct $j$. We use the statistics of $Z(u,t)$ to denote those of $Z_j(u,t)$'s.
If ${\pmb X}_0$ is a set of deterministic integers, we define $\textrm{Mom}_{Z}({\pmb X}_0)$ as the $|{\pmb X}_0|$-th moment of $Z(u,t)$ at ${\pmb X}_0$, i.e.
\begin{equation}\label{eq:cow0926}
\textrm{Mom}_{Z}({\pmb X}_0)=\textrm{E}\Biggl\{\prod_{x\in {\pmb X}_0} Z(u,x)\Biggr\}.
\end{equation}
Moreover, if $\pmb X=\{X_1,\cdots,X_n\}$ is a set consisting of random variables and ${\pmb X}_0\subseteq\pmb X$, we define $\textrm{Mom}_{Z|\pmb X}({\pmb X}_0)$ as the moment of $Z(u,t)$  conditioned on $\pmb X$, i.e.
\begin{equation}\label{eq:mouse0925}
\textrm{Mom}_{Z|\pmb X}({\pmb X}_0)=\textrm{E}\Biggl\{\prod_{x\in {\pmb X}_0} Z(u,x)\Bigr|X_1,\cdots,X_n\Biggr\}.
\end{equation}
The definitions of moments in (\ref{eq:cow0926}) and (\ref{eq:mouse0925}) hold also for random processes $Z_k(u,t)$, $z^{(N)}(u,t)$, $z_k^{(N)}(u,t)$, $z(u,t)$ and $z_k(u,t)$ depicted in Section~\ref{section:prelim}.

\begin{theorem}\label{theorem:ergodicAEM}
The unconditional AEM $\mu^{(N)}_m$, $\eta^{(N)}_{m,k}$, and $\delta^{(N)}_{m,k}$ converge a.s. to
\begin{eqnarray}
\mu_m&=&\mathop{\sum_{\varpi\in
NC(m)}}_{\varpi=\{B_1,\cdots,B_{|\varpi|}\}}\beta^{|\varpi|}\textrm{E}_{\pmb X}\left\{
\prod_{i=1}^{|\varpi|}\textrm{Mom}_{z|\pmb X}\left(
{\cal X}_{\varpi;i}\right)\right\},\label{eq:tiger0924}\\
\eta_{m,k}&=&\mathop{\sum_{\varpi\in
NC(m)}}_{\varpi=\{B_1,\cdots,B_{|\varpi|}\}} \beta^{|\varpi|-1}\textrm{E}_{\pmb X}\left\{\textrm{Mom}_{z_k|\pmb X}\left({\cal X}_{\varpi;1}\right)\prod_{i=2}^{|\varpi|}\textrm{Mom}_{z|\pmb X}\left({\cal X}_{\varpi;i}\right)\right\},\nn\\
\delta_{m,k}&=&\mathop{\sum_{\varpi\in
NC(m)}}_{\varpi=\{B_1,\cdots,B_{|\varpi|}\}}
\beta^{|\varpi|}\textrm{E}_{\pmb X}\left\{\textrm{E}_{z_k|\pmb X}(X_1)\prod_{i=1}^{|\varpi|}
\textrm{Mom}_{z|\pmb X}\left({\cal X}_{\varpi;i}\right)\right\},\nn
\end{eqnarray}
respectively, where $\pmb X=\{X_1,\cdots,X_{m-|\varpi|+1}\}$ is a set of i.i.d. random variables uniformly distributed in $[0,1)$, and ${\cal X}_{\varpi;i}=\left\{X_l: l\in\{{\cal
B}_{K(\varpi)}(t):t\in B_i\}\right\}$.
\end{theorem}
\begin{proof}
See Appendix~\ref{appendix:proof5}.
\end{proof}

Two examples of nonergodic random channel matrices are considered in the following subsections.

\subsection{Variance Profile with Each Column a Switching
Process}\label{subsection:cow0701}

Let $\bb H$ be a random matrix as described in Section~\ref{section:prelim}. Suppose that elements at the first column of $\bb H$ have the same variance (nonzero in general); for any other column $j\neq 1$, $\{V(i,j):1\leq i\leq N\}$ is a switching function from $0$ to a positive value or vice versa with a random switching time, where the positive value is random. This random matrix $\bb H$ corresponds to the channel matrix of a one-shot asynchronous CDMA to be detailed in Section~\ref{subsection:asynCDMA}.

We describe $V(i,j)$ as follows. For $j=1$,
\begin{equation}\label{eq:dragon0904}
V(i,1)=m(1),\qquad 1\leq i\leq N.
\end{equation}
For $j\in[2,K]$, $\{V(i,j):1\leq i\leq N\}$ is a switching
function from 0 to $m(j)$ or vice versa, given by
\begin{equation}\label{eq:snake0904}
V(i,j)= \left\{
\begin{array}{ll}
m(j)\textrm{u}(i-\tau(j)), & \mbox{if }w(j)=0,\\
m(j)[1-\textrm{u}(i-\tau(j))], & \mbox{if }w(j)=1,
\end{array}
\right.\quad 1\leq i\leq N,
\end{equation}
where $m(j)$ is positive and governs the magnitude, $\textrm{u}(i)$ is the unit step
function, $\tau(j)\in[0,N-1]$ specifying the switching time, and
$w(j)\in\{0,1\}$ controls the states before and after switching. Suppose that $\{m(j):1\leq j\leq K\}$, $\{\tau(j):1< j\leq K\}$ and
$\{w(j):1< j\leq K\}$ are realizations of independent random variables $M$, $T$ and $W$, respectively.
In particular, $T$ is uniform, and $W$ is equal to $0$ or $1$ equally probably.
It is clear to see the asymptotic empirical distribution of $V(1,j),V(2,j),\cdots,V(N,j)$ depends on the realizations $m(j)$, $\tau(j)$ and $w(j)$; thus, the random matrix $\bb H$ is nonergodic in one channel realization.

\begin{theorem}\label{theorem:4}
Given the random channel matrix described in (\ref{eq:dragon0904}) and (\ref{eq:snake0904}), the asymptotic $m$-th moment $\mu_m$ is
expressed as
\begin{eqnarray}\label{eq:mouse0313}
\mu_m = \mathop{\sum_{\varpi\in
NC(m)}}_{\varpi=\{B_1,\cdots,B_{|\varpi|}\}}
(\beta/2)^{|\varpi|}\prod_{i=1}^{|\varpi|} \textrm{E}\left\{M^{|B_i|}\right\}
\textrm{E}\left\{\prod_{j=1}^{|\varpi|}\left(1-\max {\cal X}_{\varpi;j}+\min {\cal
X}_{\varpi;j}\right)\right\},
\end{eqnarray}
the asymptotic $m$-th moment $\eta_{m,1}$ is
given as
\begin{eqnarray}\label{eq:mouse0902}
\eta_{m,1} =\mathop{\sum_{\varpi\in
NC(m)}}_{\varpi=\{B_1,\cdots,B_{|\varpi|}\}}
(\beta/2)^{|\varpi|-1}\prod_{i=1}^{|\varpi|} \textrm{E}\left\{M^{|B_i|}\right\}
\textrm{E}\left\{\prod_{j=2}^{|\varpi|}\left(1-\max {\cal X}_{\varpi;j}+\min {\cal
X}_{\varpi;j}\right)\right\},
\end{eqnarray}
and
\begin{equation}\label{eq:cow0902}
\delta_{m,1}=\textrm{E}\{M\}\mu_m,
\end{equation}
where ${\cal X}_{\varpi;j}$ is as defined in Theorem~\ref{theorem:ergodicAEM}.
\end{theorem}
\begin{proof}
See Appendix~\ref{appendix:4}.
\end{proof}

\subsection{Variance Profile with Each Column a Bernoulli Process}\label{subsection:grid}

Consider a random matrix $\bb H$ as described in Section~\ref{section:prelim}. Let the elements at the first column of $\bb H$ have the same variance (nonzero in general); for any other column $j\neq 1$, $\{V(i,j)\}$ is a Bernoulli process having states $m(j)>0$ and $0$ with probabilities $p(j)$ and $1-p(j)$, respectively.
In specific, when $j=1$, $V(i,1)=m(1)$; for each $j\in[2,K]$, $V(i,j)$ for $1\leq i\leq N$ are i.i.d. random variables having the density
$$
f_j(v)=(1-p(j))\delta(v)+p(j) \delta(v-m(j)),\quad 0< p(j)< 1,\quad m(j)>0,
$$
where $\{m(j):1\leq j\leq K\}$ and $\{p(j):2\leq j\leq K\}$ are realizations of independent random variables $M$ and $P$, respectively.

\begin{theorem}\label{theorem:5}
For the variance profile composed of Bernoulli processes as defined above, the
asymptotic $m$-th moment $\mu_m$ is equal to
\begin{equation}\label{eq:tiger0626}
\mu_m=\sum_{l=1}^m \beta^{l}
\mathop{\sum_{b_1+b_2+\cdots+b_l=m}}_{b_1\geq b_2\geq\cdots\geq
b_l\geq 1} \dfrac{m(m-1)\cdots(m-l+2)}{f(b_1,b_2,\cdots,b_l)}\prod_{i=1}^l
\textrm{E}\{M^{b_i}\}\textrm{E}\{P^{b_i}\},
\end{equation}
$\eta_{m,1}$ is equal to
\begin{equation}\label{eq:tiger0905}
\eta_{m,1}=\sum_{l=1}^m
\beta^{l-1}\mathop{\sum_{b_1+\cdots+b_l=m}}_{b_1\geq\cdots\geq
b_l\geq 1}\dfrac{(m-1)\cdots(m-l+2)}{f(b_1,\cdots,b_l)}
\prod_{i=1}^{l}
\textrm{E}\{ M^{b_i}\}\textrm{E}\{P^{b_i}\}
\sum_{n=1}^l \frac{b_n}{\textrm{E}\{P^{b_n}\}}.
\end{equation}
and
$$
\delta_{m,1}=\textrm{E}\{M\}\mu_m.
$$
\end{theorem}
\vspace{.2cm}
\begin{proof}
See Appendix~\ref{appendix:4}.
\end{proof}

In the following, the relation of AEM's yielded by variance profiles of switching processes and a Bernoulli process is built by some approximations. We use formulas of $\mu_m$ in (\ref{eq:mouse0313}) and (\ref{eq:tiger0626}) as an example. When the expectation
\begin{equation}\label{eq:cow0627}
\textrm{E}\left\{\prod_{j=1}^{|\varpi|}\left(1-\max {\cal
X}_{\varpi;j}+\min {\cal X}_{\varpi;j}\right)\right\}
\end{equation}
in (\ref{eq:mouse0313}) is approximated by
\begin{equation}\label{eq:tiger0627}
\prod_{j=1}^{|\varpi|}\textrm{E}\left\{1-\max {\cal
X}_{\varpi;j}+\min {\cal
X}_{\varpi;j}\right\}=\prod_{j=1}^{|\varpi|}
\left(1-\frac{|B_j|}{|B_j|+1}+\frac{1}{|B_j|+1}\right)=\prod_{j=1}^{|\varpi|}\frac{2}{|B_j|+1},
\end{equation}
we obtain (\ref{eq:mouse0313}) approximately as
\begin{eqnarray}\label{eq:mouse0627}
\mu_m&\approx&\sum_{l=1}^m \beta^{l}
\mathop{\sum_{b_1+b_2+\cdots+b_l=m}}_{b_1\geq b_2\geq\cdots\geq
b_l\geq 1} \dfrac{m(m-1)\cdots(m-l+2)}{f(b_1,b_2,\cdots,b_l)}\prod_{i=1}^l
\textrm{E}\{M^{b_i}\} (b_i+1)^{-1},
\end{eqnarray}
where the first equality of (\ref{eq:tiger0627}) holds because $X_i$'s are i.i.d. uniform random variables. Note that, when all but one of $B_i$'s are singletons, (\ref{eq:cow0627}) and
(\ref{eq:tiger0627}) are equal.

Consider the special case that, in the variance profile of Bernoulli processes,
the random variable $P$ is uniformly distributed in $(0,1)$. Then, $\textrm{E}\{P^n\}=(n+1)^{-1}$, and the asymptotic moment $\mu_m$ in (\ref{eq:tiger0626}) becomes the same as
(\ref{eq:mouse0627}). The coincidence of the two expressions can be interpreted as follows. For the variance profile as a switching process, each column of $\bb H$ has a continuous segment of zeros with its length uniformly distributed in $[0,N)$. If these zeros are spread out randomly to the whole interval $[0,N)$, the variance profile of switching processes becomes a variance profile of Bernoulli processes.

\section{Applications}

In this section, two applications of the AEM formulas derived above are provided. In the first subsection, the spectral efficiency of an MIMO channel given a conditional variance profile is computed.
Our intention is to exemplify the use of the moment based method, instead of the widespread Stieltjes transform based approach, in determining the spectral efficiency of a communication system. In the second subsection, we consider a one-shot symbol-asynchronous but chip-synchronous CDMA system, where the random variance profile depicted in Section~\ref{subsection:cow0701} (as well as \ref{subsection:grid}, approximately) is the corresponding variance profile.

\subsection{Correlated Fading in MIMO}

In some applications of probability, it is frequent that the
(infinite) moment sequence of an unknown distribution $F$ is
available, and these moments determine a unique distribution.
Suppose that the final aim is to calculate the expected value of
function $g(X)$ of the random variable $X$ whose distribution $F$ is
unknown. One of the most widely used techniques for evaluating
$\textrm{E}\{g(X)\}$ is the Gauss quadrature method
\cite{golub69}, where $2Q+1$ moments $\{m_n\}_{n=0}^{2Q}$ of
distribution $F$ are used to determine a $Q$-point quadrature rule
$\{w_q,x_q\}_{q=1}^Q$ such that
\begin{equation}\label{eq:mouse0905}
\textrm{E}\{g(X)\}=\int_{-\infty}^\infty g(x)\textrm{d}F(x)\approx \sum_{q=1}^Q w_q g(x_q),
\end{equation}
and the approximation error becomes negligible when $Q$ is large. In this subsection, the Gauss quadrature method is used to compute the mutual information of a spatially correlated MIMO system.

Consider $n_\textrm{T}$ transmit and $n_\textrm{R}$ receive antennas
with the corresponding multiantenna channel denoted by $\bb H$,
whose $(i,j)$-th entry is the fading coefficient between
the $j$-th transmit antenna and $i$-th receive antenna. Let the
correlation between the $(i,j)$-th and $(i',j')$-th entries of $\bb
H$ be represented by
$$
R_{\bb H}(i,j;i',j')=\textrm{E}\{H_{i,j}H_{i',j'}^*\}.
$$
It is shown in \cite{tulino05_1} that any MIMO channel with
correlation $R_{\bb H}$ that falls within the
unitary-independent-unitary (UIU)\footnote{For the definition of
UIU, see \cite{tulino05_1}} framework and has bounded eigenvalues is
unitarily equivalent to an i.n.d. channel $\tilde{\bb H}$ with the mean and variance of the
$(i,j)$-th entry equal to $0$ and the $(i,j)$-th eigenvalue of $R_{\bb
H}$, respectively. That is, $\bb H\bb H^\dag$ and $\tilde{\bb H}\tilde{\bb H}^\dag$ have the same AED.

We consider Example 2 of \cite{tulino05_1}, where $n_\textrm{T}=3$,
$n_\textrm{R}=2$, and an i.n.d. channel $\tilde{\bb H}$ whose component is zero-mean and has variance equal to the entry of $\bb G$ at the same location with
\begin{equation}\label{eq:cow0904}
\bb G=\left[
\begin{array}{ccc}
0.4 & 3.6 & 0.5\\
0.3 & 1 & 0.2
\end{array}
\right];
\end{equation}
each $(i,j)$-th entry of $\tilde{\bb H}$ has independent real and
imaginary parts uniformly distributed in the interval
$[-\sqrt{1.5 G_{i,j}},\sqrt{1.5 G_{i,j}}]$. It is known that the normalized input-output mutual information is
$$
\frac{1}{n_\textrm{R}}\sum_{i=1}^{n_\textrm{R}}\log_2\left(1+\gamma\lambda_i\left(\tilde{\bb H}\bb\Phi\tilde{\bb H}^\dag\right)\right)\mathop{\longrightarrow}^{\textrm{a.s.}}\int\log_2(1+\gamma X)\textrm{d}F(X),
$$
where $\gamma$ is the input SNR, $\lambda_i(\cdot)$ is the $i$-th eigenvalue of the indicated matrix, $X$ is the limiting random variable governing the eigenvalue distribution of $\tilde{\bb H}\bb\Phi\tilde{\bb H}^\dag$ with $\bb \Phi$ the diagonal input covariance matrix, and $F(X)$ is the cumulative distribution function of $X$. Thus, the liming mutual information can be obtained by letting $g(x)=\log_2(1+\gamma x)$ in (\ref{eq:mouse0905}), and the AEM of $\tilde{\bb H}\bb\Phi\tilde{\bb H}^\dag$ can be acquired from those of $\tilde{\bb H}\tilde{\bb H}^\dag$ and $\bb\Phi$ using the result of \cite{yin83}.

\begin{figure}[t]
\begin{center}
\begin{tabular}{c}
\psfig{figure=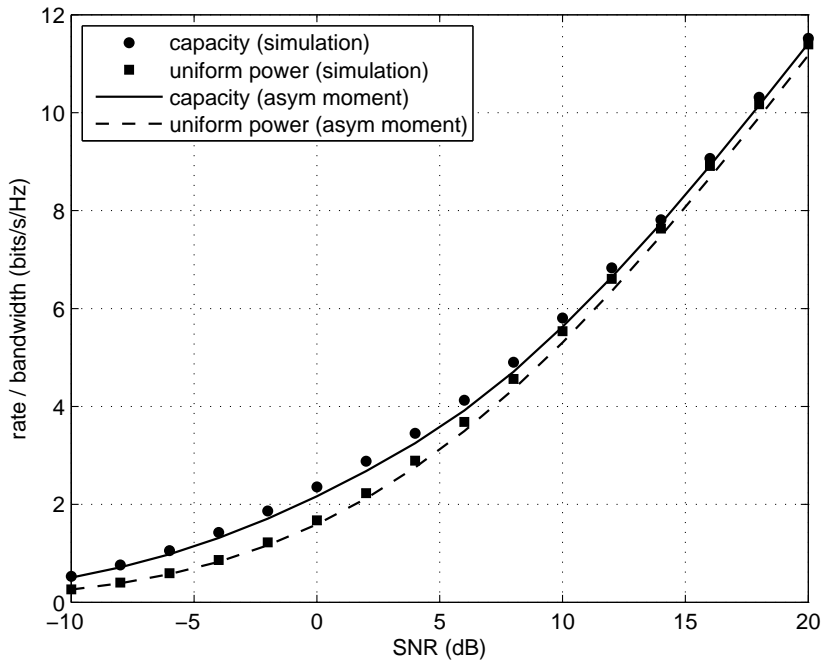,height=9cm}
\end{tabular}
\end{center}
\caption{The capacity and the spectral efficiency with uniform power allocation of a spatially correlated MIMO with variance profile given in (\ref{eq:cow0904}). The two analytical curves (solid and dashed lines) are obtained by 5-point Gauss quadrature method.} \label{fig:mouse0904}
\end{figure}

Fig.~\ref{fig:mouse0904} shows both the simulated and analytical results of the capacity and the spectral efficiency with uniform power allocation versus input SNR, where the capacity-achieving power allocation $\bb\Phi$ is found using the algorithm in \cite{tulino06}, and two curves showing analytical results (solid and dashed lines) are obtained using a 5-point Gauss quadrature rule. The analytical results in Fig.~\ref{fig:mouse0904} using the moment-based method have a slight performance gain compared with the transform-based method presented in \cite{tulino05_1}. The gain in terms of error percentage\footnote{The error percentage is defined as (simulation result-analytical result)/simulation result $\times$ 100\%.} becomes more and more obvious when the input SNR gets larger. When the input SNR is equal to 20dB, capacities based on the Monte-Carlo simulation, the methods in \cite{tulino05_1} and this paper are $11.49$, $11.09$, and $11.41$ (bits/s/Hz), respectively; spectral efficiency with uniform power of the three methods
in the same order are $11.37$, $10.99$, and $11.17$ (bits/s/Hz).

\subsection{One-Shot Asynchronous DS-CDMA}\label{subsection:asynCDMA}

Consider a DS-CDMA system with asynchronous transmission where each user's spreading
sequence is chosen randomly and independently. It is assumed that
the system is chip-synchronous, i.e., the relative delay of each
user is an integer multiple of the chip duration.

Suppose that $b_k\in\{+1,-1\}$ is the transmitted symbol of user
$k$, $\bb s_k\in\{+1/\sqrt{N},-1/\sqrt{N}\}^N$ is the signature sequence of the same
user, and $\bb n$ is the AWGN with variance $\sigma^2$.
We adopt the signal model analogous to that presented in
[\citenum{kiran00}, eqn. (2)]; the sampled discrete-time model for
the received signal is given as
\begin{equation}\label{eq:mouse0630}
\bb r=\alpha_1 b_1 \bb s_1+\sum_{k=2}^K \alpha_k b_k \bb
u_k+\sum_{k=2}^K \alpha_k d_k \bb v_k +\bb n,
\end{equation}
where $\alpha_k$ is the fading coefficient of the $k$-th user,
$b_k,d_k\in{\mathbb R}$ are two consecutive symbols of the $k$-th
user which overlap with user $1$ in the observation window. Symbols
$b_k$ and $d_k$ have effective signature sequences $\bb
u_k$ and $\bb v_k$, respectively. If
$t_k\in{\mathbb Z}^+$ denotes the relative delay of user $k$ to user $1$ in terms of the number
of chips, then
$$
(\bb u_k)_i=\left\{
\begin{array}{cc}
{(\bb s_k)}_{N-t_k+i},& 1\leq i\leq t_k\\
0, & t_k<i\leq N
\end{array}
\right.\quad\mbox{and}\quad (\bb v_k)_i=\left\{
\begin{array}{cc}
0, & 1\leq i\leq t_k\\
{(\bb s_k)}_{i-t_k},& t_k<i\leq N
\end{array}
\right.
$$
The signal $\bb r$ in (\ref{eq:mouse0630}) can be expressed in a
more compact form as
\begin{equation}\label{eq:mouse0701}
\bb r=\bb S\bb A\bb b+\bb n,
\end{equation}
where the notations are defined as follows:
\begin{eqnarray}
&& \bb S=[\bb u_1,\bb u_2,\cdots,\bb u_K,\bb v_2,\cdots,\bb
v_K]\in{\mathbb R}^{N\times(2K-1)},\nn\\
&& \bb A=\diag(\alpha_1,\alpha_2,\cdots,\alpha_K,\alpha_2,\cdots,\alpha_K)\in{\mathbb C}^{(2K-1)\times(2K-1)},\nn\\
&& \bb b = [b_1,b_2,\cdots,b_K,d_2,\cdots,d_K]^T\in{\mathbb
R}^{(2K-1)\times 1}.\nn
\end{eqnarray}
Define $\bb H=\bb S\bb A$ and $\bb R=\bb H^\dag\bb H$. Following the
methodology of \cite{moshavi96,cottatellucci05_1}, we obtain a $D$-dimensional
reduced-rank MMSE receiver for user 1, i.e. the estimate of $b_1$
as $\hat{b}_1(\bb r)=\bb w^T \bb r$, where
\begin{equation}\label{eq:dragon0905}
\bb w=\bb\Phi^{-1}\pmb\varphi,
\end{equation}
with
\begin{equation}\label{eq:snake0905}
\pmb\varphi=[(\bb R)_{11},(\bb R^2)_{11},\cdots,(\bb R^D)_{11}]^T
\quad\mbox{and}\quad(\bb\Phi)_{ij}=(\bb R^{i+j})_{11}+\sigma^2(\bb
R^{i+j-1})_{11}, \quad 1\leq i,j\leq D.
\end{equation}
It is seen $(\bb R^m)_{11}$ is equal to $\eta_{m,1}^{(N)}$ defined in (\ref{eq:cow0427}).

Suppose that the relative delays $t_2,\cdots,t_K$ are realizations of a discrete random variable uniformly distributed in $[0,N)$. The random matrix $\bb H=\bb S\bb A$ has a conditional variance profile identical to that presented in Section~\ref{subsection:cow0701}, where $|\alpha_1|^2,\cdots,|\alpha_K|^2$ are realizations of random variable $M$. The AEM formula $\eta_{m,1}$ given in (\ref{eq:mouse0902}) can be used to obtain the reduced-rank MMSE receiver $\bb w=\bb\Phi^{-1}\pmb\varphi$ in the asymptotic regime. However, the computation of $\eta_{m,1}$ is involved with several multi-dimensional integrals, and the complexity is quite high. We may resort to its counterpart in (\ref{eq:tiger0905}) whose variance profile is composed of Bernoulli processes, and the random variable $P$ therein is set as uniform in $(0,1)$. As demonstrated in the discussion following Theorem~\ref{theorem:5}, the use of (\ref{eq:tiger0905}) as a replacement of (\ref{eq:mouse0902}) is an approximation by spreading the zero elements in each column of channel matrix $\bb H$ randomly and uniformly.

Simulations are run to compare the AEM's by (\ref{eq:mouse0902}) and (\ref{eq:tiger0905}). The AEM of the former are obtained through $10^5$-point Monte-Carlo integration, and we evaluate the first eight moments for various values of $\beta$. Since the distinction of the two sets of AEM is on the distribution of zero entries in columns of $\bb H$ and irrelevant to the fading process $M$, we let $M$ as a constant equal to $1$. The random variable $P$ in (\ref{eq:tiger0905}) is set uniform in $(0,1)$. The results show the error percentage\footnote{The error percentage is defined as $\dfrac{\mbox{(\ref{eq:mouse0902})-(\ref{eq:tiger0905})}}{\mbox{(\ref{eq:mouse0902})}}$ $\times$ 100\%.} increases with the order $m$ of moments, equal to $0.02$\%, $0.22$\%, $0.90$\% and $1.78$\% for $m=2$, $4$, $6$ and $8$, respectively. It is seen that AEM obtained from (\ref{eq:mouse0902}) have larger values than those from (\ref{eq:tiger0905}).

\begin{figure}[t]
\begin{center}
\begin{tabular}{c}
\psfig{figure=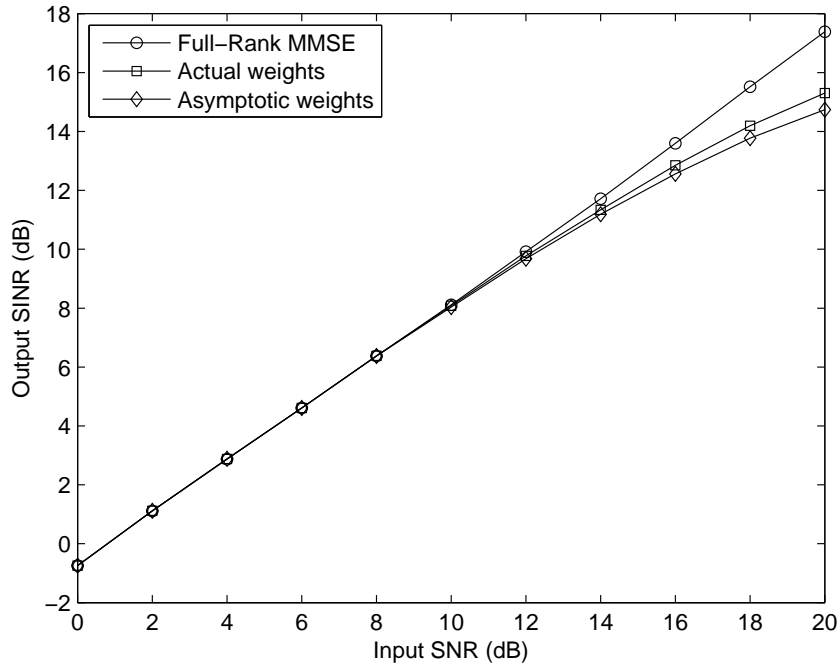,height=9cm}
\end{tabular}
\end{center}
\caption{The output SINR of three receivers in a symbol-asynchronous but chip-synchronous CDMA system with an unfaded channel: $K=16$, $N=64$, and $D=4$.} \label{fig:cow0904}
\end{figure}

\begin{figure}[t]
\begin{center}
\begin{tabular}{c}
\psfig{figure=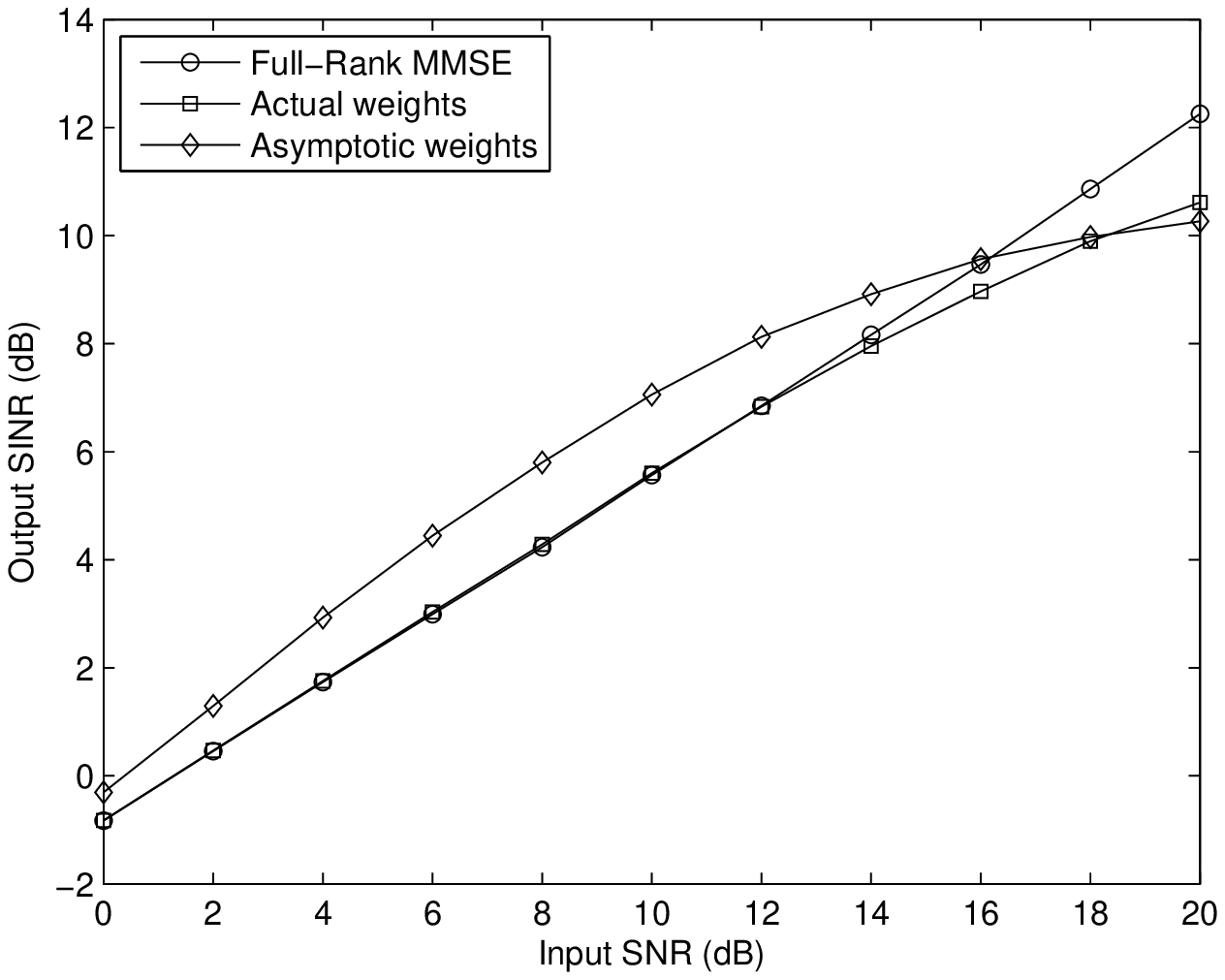,height=9cm}
\end{tabular}
\end{center}
\caption{The output SINR of three receivers in a symbol-asynchronous but chip-synchronous CDMA system with a Rayleigh flat fading channel: $K=16$, $N=64$, and $D=4$.} \label{fig:tiger0904}
\end{figure}

\begin{figure}[h]
\begin{center}
\begin{tabular}{c}
\psfig{figure=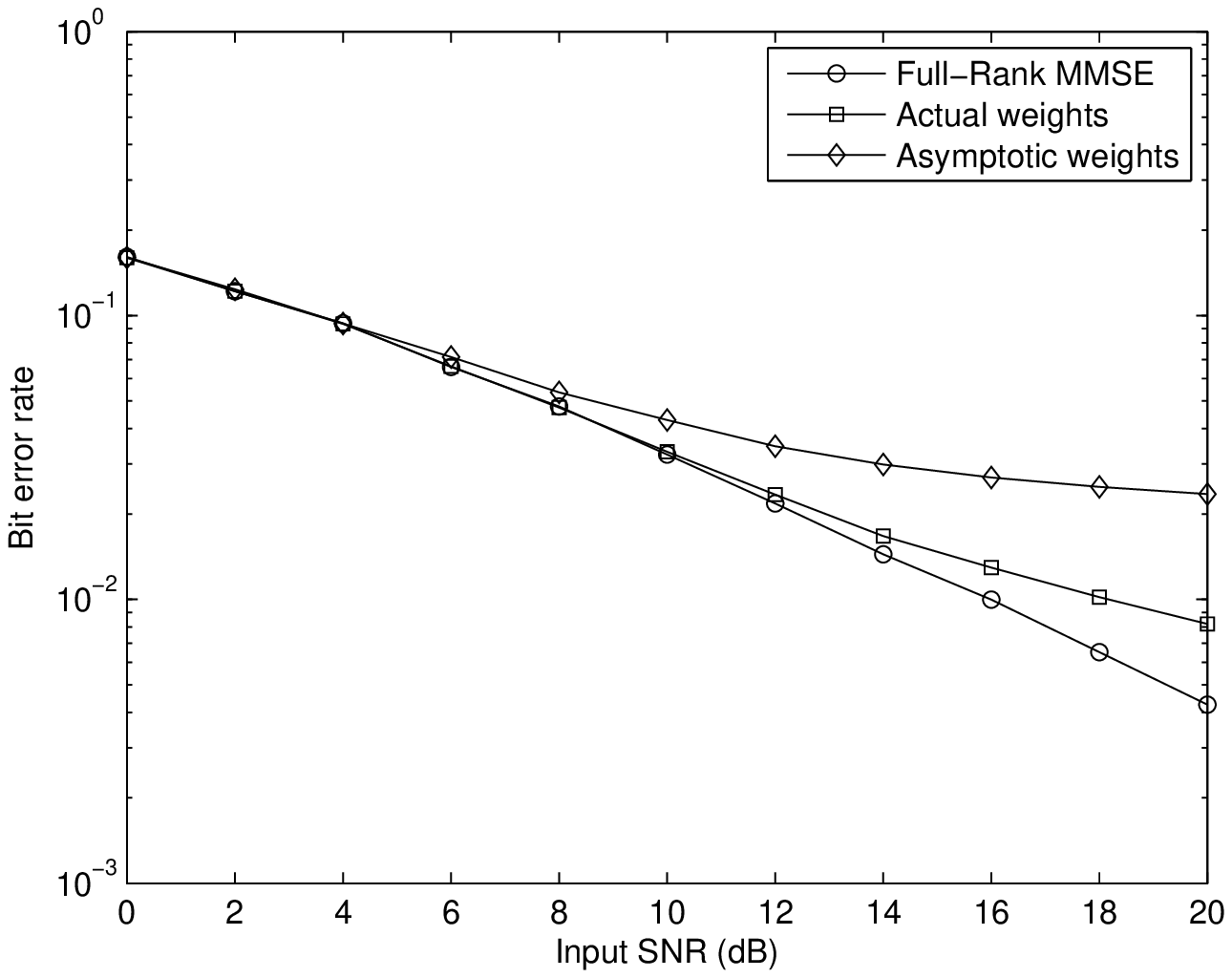,height=9cm}
\end{tabular}
\end{center}
\caption{The BER of three receivers in a symbol-asynchronous but chip-synchronous CDMA system with a Rayleigh flat fading channel: $K=16$, $N=64$, and $D=4$.} \label{fig:rabbit0904}
\end{figure}

In the following, we compare the output SINR of three receivers: i) the full-rank MMSE receiver, ii) the reduced-rank MMSE receiver that uses the actual weights, and iii) the asymptotic reduced-rank MMSE receiver that uses the asymptotic weights. For the second receiver, the filter in (\ref{eq:dragon0905}) is employed; while for the third receiver, entries of $\bb\Phi$ and $\pmb\varphi$ in (\ref{eq:snake0905}) are replaced with their limiting values in the asymptotic regime. The number of users is $K=16$, and the spreading gain is $N=64$. The two receivers have the rank $D=4$. For the asymptotic weights of receiver iii), we use (\ref{eq:tiger0905}) to compute $\eta_{m,1}$ because it yields a very close result to (\ref{eq:mouse0902}) and it demands lower complexity. The output SINR is the ratio of total signal power and total interference-plus-noise power of a large number of independent simulations.

Fig.~\ref{fig:cow0904} shows the output SINR for the three receivers in an unfaded channel. It is seen that the output SINR of the asymptotic reduced-rank receiver is close to that of the receiver using actual weights for the input SNR considered, although the asymptotic weights of the third receiver are obtained using the ensemble statistics of relative delays and spreading codes.

Fig.~\ref{fig:tiger0904} plots the performance of the three receivers in a Rayleigh flat-fading channel. In formula (\ref{eq:tiger0905}), the moments of $M$ are obtained as those of a complex Gaussian with independent real and imaginary parts. As the empirical distribution of the fading coefficients $\alpha_1,\cdots,\alpha_K$ can hardly approach Gaussian in one channel realization, the asymptotic weights can be viewed as calculated using a fading statistic of a very long observation time. It is seen from the figure that the output SINR of the asymptotic reduced-rank receiver is larger than the other two receivers until the input SNR exceeds $16$dB. This is because, for the asymptotic receiver, fading coefficients of extremely large magnitudes that occur occasionally result in profound effects on the output signal power. For the other two receivers, the effect of fading coefficients with large magnitudes can be offset by the MMSE filtering that is computed from the known fading coefficients, while the asymptotic receiver fails to do so because its weights are computed based on the ensemble fading process over a long time. In Fig.~\ref{fig:rabbit0904}, we show the bit error rates (BER) versus input SNR of the three receivers in the Rayleigh fading environments. Although the asymptotic reduced-rank receiver has the largest output SINR, its BER performance is worse than the other two receivers as expected. To boost the performance of the asymptotic receiver, we can employ the implementation in \cite{li04} that the moments of the fading process are found by the empirical distribution of finite number of realizations.

\section{Conclusions}

In this paper, we have used noncrossing partition to derive expressions of several AEM related quantities of a large Wishart-type random matrix $\bb H\bb H^\dag$ when the variance profile of $\bb H$ is random and $\bb H$ is nonergodic in one channel realization. These quantities are useful in the design and analysis of a number of communication systems. It was assumed that $\bb H=\bb M\odot\bb S$, where entries of $\bb S$ are zero-mean i.i.d. random variables, and columns $\{|M_{i,j}|^2:1\leq i\leq N\}$ are realizations of i.i.d random processes. The derivation started from obtaining the conditional AEM given a certain realization of $\bb M$, and we then derived unconditional AEM by computing ensemble average of the conditional quantity according to the distribution of $\bb M$. We have obtained conditional AEM expressions in both direct and recursive forms. When the conditional variance profile is decomposable as the product of two one-dimensional functions with deterministic nonnegative values, the conditional AEM expressions can be simplified. Two statistical models of $\bb M$ were given. One is that each column $\{|M_{i,j}|^2:1\leq i\leq N\}$ is a switching function with a random switching time; the other is that $\{|M_{i,j}|^2:1\leq i\leq N\}$ is a Bernoulli process with a random success probability.

Two application cases have been provided. One is the use of Gauss quadrature method to computing the ergodic capacity of a spatially correlated MIMO system by means of AEM; a better accuracy than the Stieltjes transform based method of \cite{tulino05_1} has been achieved. The other is the design of a reduced-rank MMSE receiver in a one-shot asynchronous CDMA system, whose corresponding variance profile has each column as a switching function with random switching time; the AEM calculation is involved with multidimensional integrals. Due to the difficulty in evaluating the integrals, a variance profile consisting of Bernoulli processes was used instead. Numerical results showed the approximation error is negligible.

\appendices

\section{$K$-graph}\label{appendix:K-graph}

\begin{figure}[t]
\begin{center}
\begin{tabular}{c}
\psfig{figure=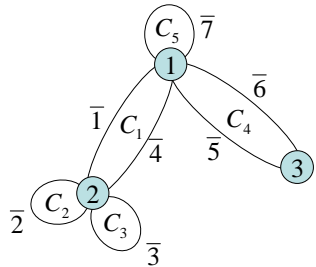,height=3.8cm}
\end{tabular}
\end{center}
\caption{The $K$-graph of noncrossing partition
$\varpi=\{\{1,5,7\},\{2,3,4\},\{6\}\}$, which is composed of five
cycles $C_1,\cdots,C_5$.} \label{fig:mouse0411}
\end{figure}

\begin{definition}[$K$-graph \cite{hwang06}]\label{definition:4}
The $K$-graph of a noncrossing partition
$\varpi=\{B_1,\cdots,B_j\}\in NC(m)$ is denoted by $G=({\cal
V},{\cal E})$. The vertex set is ${\cal V}=\{1,2,\cdots,j\}$, and
the edge set is ${\cal
E}=\{\overline{1},\overline{2},\cdots,\overline{m}\}$. Edge
$\overline{r}$ connects vertices $s$ and $t$ if ${\cal
B}_\varpi(r)=s$ and ${\cal B}_\varpi(r+1)=t$ (with $m+1:=1$).
$\hfill{\small \blacksquare}$
\end{definition}
The $K$-graph of $\varpi=\{\{1,5,7\},\{2,3,4\},\{6\}\}\in NC(7)$ is
shown in Fig.~\ref{fig:mouse0411}. The properties of a $K$-graph are
summarized below.
\begin{enumerate}
\item A noncrossing partition can be recovered from its $K$-graph. Blocks
of $\varpi$ can be identified by starting at vertex $1$ and
traversing edges $\overline{1},\overline{2},\cdots,\overline{m}$; if
edge $\overline{r}$ starts at vertex $i$, then $r\in B_i$. For
example, in Fig.~\ref{fig:mouse0411}, edges
$(\overline{1},\overline{2},\cdots,\overline{7})$ start at vertices
$(1,2,2,2,1,3,1)$, respectively.

\item Due to the noncrossing property, a $K$-graph of $\varpi\in
NC(m)$ and $|\varpi|=j$ is a concatenation of $m-j+1$ cycles with
any pair of cycles connecting by at most one vertex. The $K$-graph
in Fig.~\ref{fig:mouse0411} is a concatenation of $7-3+1=5$ cycles,
denoted by $C_1,\cdots,C_5$. The cycles are ordered in the ascending
order of the minimum edge element in them.

\item The $i$-th block of $K(\varpi)$ is recognized by the edges in
the $i$-th cycle of the $K$-graph. In Fig.~\ref{fig:mouse0411},
since edges $\{\overline{1}, \overline{4}\}$, $\{\overline{2}\}$,
$\{\overline{3}\}$, $\{\overline{5},\overline{6}\}$, and
$\{\overline{7}\}$ constitute cycles $C_1, C_2, C_3,C_4$ and $C_5$,
respectively, we have
$K(\varpi)=\{\{\overline{1},\overline{4}\},\{\overline{2}\},
\{\overline{3}\}, \{\overline{5},\overline{6}\},\{\overline{7}\}\}$.

\item There is a bijective correspondence between blocks of a noncrossing partition $\varpi$
and vertices of its $K$-graph. There is a bijective correspondence
between blocks of the Kreweras complementation map $K(\varpi)$ and
cycles of the $K$-graph of $\varpi$.
\end{enumerate}

\textit{Remark}: A $K$-graph can be interpreted in a more visually
convenient way as follows. Let us arrange vertices $v_1,\cdots,v_m$
orderly (either clockwise or counter-clockwise) in an $m$-vertex
cycle, and let edge $\overline{r}$ connect vertices $v_r$ and
$v_{r+1}$. The $K$-graph of $\varpi\in NC(m)$ can be obtained by,
for $1\leq i\leq |\varpi|$, merging vertices $\{v_j:j\in B_i\}$ into
a vertex $i$. When $v_j$'s are merged into a new one, edges
originally incident on $v_j$'s become incident on the new vertex.
Mergence of two adjacent vertices results in a self-loop cycle.
\hfill{\small $\blacksquare$}

Take Fig.~\ref{fig:mouse0411} as an example. Vertex $1$ is formed by
the mergence of vertices $v_1,v_5$ and $v_7$ of a $7$-vertex cycle.
Edges $\overline{1}$ and $\overline{7}$, $\overline{4}$ and
$\overline{5}$, and $\overline{6}$ and $\overline{7}$ originally
incident on vertices $v_1$, $v_5$ and $v_7$, respectively, become
incident on the new vertex $1$ after mergence. Moreover, since
vertices $v_1$ and $v_7$ are adjacent in a 7-vertex cycle, their mergence results in a
self-loop cycle $C_5$.

\section{Proof of Theorem~\ref{theorem:1}}\label{appendix:proof1}

\subsection{Proof for $\hat{\mu}_m$}

For the ease of extending the derivation of $\hat{\mu}_m$ to $\hat{\eta}_{m,k}$ and $\hat{\delta}_{m,k}$,
we rewrite the right-hand-side of (\ref{eq:mouse0427}) as
\begin{equation}\label{eq:mouse0712}
\beta\cdot K^{-1}\tr\{(\bb H^\dag\bb H)^m\},
\end{equation}
and the operand of the limit in the expression of $\hat{\mu}_m$ is given as $\beta K^{-1}\textrm{E}\{\tr\{(\bb H^\dag\bb H)^m\}|v^{(N)}(x,y)\}$. For simplicity, the conditional notation $\{\cdot|v^{(N)}(x,y)\}$ is omitted below. We have
\begin{eqnarray}
&&K^{-1} \textrm{E}\{\tr\{(\bb H^\dag\bb H)^m\}\}\nn\\
&=& K^{-1}\sum_{\{k_1,\cdots,k_m\}\in[1,K]^m}\textrm{E}\{(\bb
H^\dag\bb
H)_{k_1 k_2}(\bb H^\dag\bb H)_{k_2 k_3}\cdots (\bb H^\dag\bb H)_{k_m k_1}\}\nn\\
&=&
K^{-1}\mathop{\sum_{\{k_1,\cdots,k_m\}\in[1,K]^m}}_{\{n_1,\cdots,n_m\}\in[1,N]^m}
\textrm{E}\left\{\left(H_{n_m,k_1}^* H_{n_1,k_1}\right)
\left(H_{n_1,k_2}^* H_{n_2,k_2}\right)\cdots\left(H_{n_{m-1},k_m}^*
H_{n_m,k_m}\right)\right\}.\label{eq:mouse0412}
\end{eqnarray}
To evaluate the expectation in (\ref{eq:mouse0412}), it is required
to consider the equivalence relation of variables $k_1,\cdots,k_m$
as well as that of $n_1,\cdots,n_m$. As the equivalence relation is
equivalent to set partitioning by that \textit{two variables take the
same value (in the range of $[1,K]$ or $[1,N]$ depending on whether
$k_i$'s or $n_i$'s are considered) if and only if they are
partitioned in the same block}, set partition theory can be employed
as the tool for expectation evaluation of (\ref{eq:mouse0412}).
Thus, we investigate partitions of two totally ordered set
$\{k_1,\cdots,k_m\}$ and $\{n_1,\cdots,n_m\}$. It is known
that, when evaluating (\ref{eq:mouse0412}) asymptotically, it is
sufficient to consider the situations that $k_1,\cdots,k_m$ is
partitioned noncrossingly. We will show that, to yield non-vanishing expectation of (\ref{eq:mouse0412}),
the partition of $\{n_1,\cdots,n_m\}$ should be the Kreweras
complementation map of the partition of $\{k_1,\cdots,k_m\}$. In the
following, for convenience, we denote the partitions of
$\{k_1,k_2,\cdots,k_m\}$ and $\{n_1,n_2,\cdots,n_m\}$ by those of
$\{1,2,\cdots,m\}$ and
$\{\overline{1},\overline{2},\cdots,\overline{m}\}$, respectively.

\begin{figure}[t]
\begin{center}
\begin{tabular}{c}
\psfig{figure=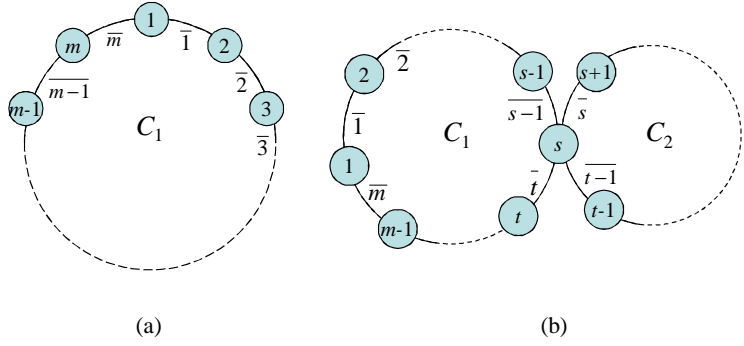,height=5cm}
\end{tabular}
\end{center}
\caption{The $K$-graphs of (a) $m$-block noncrossing partition $\{\{1\},\{2\},\cdots,\{m\}\}$, and (b) $(m-1)$-block noncrossing partition $\{\{1\},\{2\},\cdots,\{s,t\},\{s+1\},\cdots\{m\}\}$ with
$s<t$.} \label{fig:rabbit0412}
\end{figure}

We consider $\varpi=\{B_1,\cdots,B_j\}\in NC(\{1,\cdots,m\})$ for
various values of $j$. First, we consider $j=m$, resulting in
$\varpi=\{\{1\},\{2\},\cdots,\{m\}\}$. By the connection of
equivalence relation and set partitioning, this corresponds to $k_1\neq
\cdots\neq k_m$ in (\ref{eq:mouse0412}). In this case, since entries
of $\bb H$ are independent, (\ref{eq:mouse0412}) becomes
\begin{eqnarray}\label{eq:cow0412}
&&K^{-1}\mathop{\sum_{\textrm{all $k_i$'s
distinct}}}_{n_1,\cdots,n_m} \textrm{E}\{H_{n_m,k_1}^* H_{n_1,k_1}\}
\textrm{E}\{H_{n_1,k_2}^*
H_{n_2,k_2}\} \cdots\textrm{E}\{H_{n_{m-1},k_m}^* H_{n_m,k_m}\}.
\end{eqnarray}
To yield a nonzero summand in (\ref{eq:cow0412}), it is required
that $n_1=\cdots=n_m$. Interpreting this equivalence relation by
partition of $\{n_1,\cdots,n_m\}$, we can see it corresponds to
the Kreweras complementation map of $\{\{1\},\cdots,\{m\}\}$, i.e.
$K(\{\{1\},\cdots,\{m\}\})=\{\{\overline{1},\cdots,\overline{m}\}\}$.
The sum in (\ref{eq:cow0412}) thus becomes
\begin{equation}\label{eq:tiger0412}
K^{-1}\mathop{\sum_{\{b_1,\cdots,b_m\}\in[1,K]^m}}_{b_1\neq\cdots\neq
b_m}\sum_{c_1\in[1,N]}\prod_{i=1}^m \dfrac{V(c_1,b_i)}{N}.
\end{equation}
The $K$-graph of $\{\{1\},\cdots,\{m\}\}$ is shown in
Fig.~\ref{fig:rabbit0412}(a), where there are $m$ vertices
$1,\cdots,m$, and the $K$-graph is itself a cycle labelled as $C_1$.
The variable $b_i$ in (\ref{eq:tiger0412}) is interpreted as the
integer in $[1,K]$ chosen by the $i$-th block of the partition of
$\{\{1\},\cdots,\{m\}\}$ (or vertex $i$ of the $K$-graph), and $c_1$
in (\ref{eq:tiger0412}) is the integer chosen by the block of
$\{\{\overline{1},\cdots,\overline{m}\}\}$ (or the cycle of the
$K$-graph). In the sequel, $b_i$ (or $c_j$) is used to represent the
integer in $[1,K]$ (or $[1,N]$) selected by vertex $i$ (or cycle
$C_j$).

Next, consider $\varpi=\{B_1,\cdots,B_j\}\in NC(\{1,\cdots,m\})$
with $j=m-1$, where one of the blocks contains two elements $s$ and
$t$ ($s<t$), and all other blocks are singletons. In this case,
(\ref{eq:mouse0412}) becomes
\begin{eqnarray}\label{eq:mouse0416}
&&K^{-1}\mathop{\sum_{\textrm{all $k_i$'s distinct except for
$k_s=k_t$}}}_{n_1,\cdots,n_m} \textrm{E}\{H_{n_m,k_1}^*
H_{n_1,k_1}\}
\textrm{E}\{H_{n_1,k_2}^* H_{n_2,k_2}\} \cdots\nn\\
&&\textrm{E}\{H_{n_{s-1},k_s}^* H_{n_s,k_s} H_{n_{t-1},k_t}^*
H_{n_t,k_t}\} \cdots \textrm{E}\{H_{n_{m-1},k_m}^* H_{n_m,k_m}\}.
\end{eqnarray}
To yield a nonzero summand of (\ref{eq:mouse0416}), we need that the
variables $n_1,\cdots,n_m$ in each expectation of
(\ref{eq:mouse0416}) are paired. That is, $n_m=n_1$ in the first
expectation, $n_1=n_2$ in the second, and so on. In the expectation
containing $n_{s-1},n_s,n_{t-1}$ and $n_t$, there are three possible
cases for pairing, i.e., (i) $n_{s-1}=n_s$ and $n_{t-1}=n_t$, (ii)
$n_s=n_t$ and $n_{s-1}=n_{t-1}$, and (iii) $n_{s-1}=n_t$ and
$n_{t-1}=n_s$. Among these, case (iii) (along with $n_m=n_1$,
$n_1=n_2$, and so on) divides $n_1,\cdots,n_m$ into two groups
$(n_1,\cdots,n_{s-1},n_t,\cdots,n_m)$ and $(n_s,\cdots,n_{t-1})$
with $n_i$'s in the same group take the same value. On the contrary,
cases (i) and (ii) result in only one group. Thus, when computing
(\ref{eq:mouse0416}) in the limit $N\to\infty$, it is sufficient
to consider only case (iii) since it yields the highest dimension of
$N$. Note that the grouping of $n_i$'s resulting from case (iii) can
be obtained from
$K(\{\{1\},\{2\},\cdots,\{s,t\},\cdots\{m\}\})=\{\{\overline{1},\cdots,\overline{s-1},\overline{t},
\cdots,\overline{m}\},\{\overline{s},\cdots,\overline{t-1}\}\}$. The
$K$-graph corresponding to the current case is shown in
Fig.~\ref{fig:rabbit0412}(b). We notice that case (iii) is
equivalent to letting edges in the same cycle be in the same
group. By assigning $b_i$ and $c_j$ to the $i$-th vertex and the
$j$-th cycle, respectively, (\ref{eq:mouse0416}) is given
by
\begin{eqnarray}\label{eq:mouse0417}
&&K^{-1}\mathop{\sum_{b_1,\cdots,b_{m-1}}}_{b_1\neq \cdots\neq
b_{m-1}}\mathop{\sum_{c_1,c_2}}_{c_1\neq c_2} \frac{V(c_1,b_1)}{N}
\frac{V(c_1,b_2)}{N} \cdots \frac{V(c_1,b_s)V(c_2,b_s)}{N^2}\cdots
\frac{V(c_1,b_{m-1})}{N},
\end{eqnarray}
where $V(c_1,b_1)/N$, $V(c_1,b_2)/N$ and $V(c_1,b_{m-1})/N$ are
equal to the first, second and fourth expectations of
(\ref{eq:mouse0416}), respectively, while $V(c_1,b_s)V(c_2,b_s)/N^2$
comes from the third expectation. The expression of (\ref{eq:mouse0417})
can be recovered directly from Fig.~\ref{fig:rabbit0412}(b). In
specific, each vertex $i$ is associated with
$$
\prod_{j\in T_i}\frac{V(c_j,b_i)}{N},
$$
where
$j\in T_i$ if vertex $i$ resides in cycle $C_j$, and (\ref{eq:mouse0417}) is
obtained by multiplying the terms corresponding to each vertex
altogether. For example, in Fig.~\ref{fig:rabbit0412}(b), vertex $s$
resides in cycles $C_1$ and $C_2$, while vertex $1$ resides only in
$C_1$. Thus, $T_1=\{1\}$ and $T_s=\{1,2\}$. We can obtain (\ref{eq:tiger0412}) from Fig.~\ref{fig:rabbit0412}(a) in the same manner.

In the above, we consider noncrossing partitions with $m$ and $m-1$ blocks. Here we summarize the findings in the two cases and extend to noncrossing partitions with arbitrary number of blocks.
According to the remark following Definition~\ref{definition:4}, the
$K$-graph of any noncrossing partition $\varpi$ of
$\{1,\cdots,m\}$ can be obtained by successively merging two
vertices of a same cycle.\footnote{If two vertices of distinct cycles are merged,
it results in a crossing partition.} This vertex mergence starts from an $m$-vertex cycle (Fig.~\ref{fig:rabbit0412}(a)) where all edges are assigned with an identical integer in $[1,N]$. At the first iteration, two vertices of the $m$-vertex cycle are merged into one, yielding a concatenation of two cycles (Fig.~\ref{fig:rabbit0412}(b)). Edges in this two-cycle graph take the same integer if and only if they are in the same cycle. At the $r$-th iteration, $1\leq r\leq m-|\varpi|$, two vertices $s$ and $t$ in any one of the $r$ cycles
are merged to yield $r+1$ cycles in total. The cycle in which vertices $s$ and $t$ originally locate is then torn into two, and the edges in these two cycles are assigned with the same integer if and only if they are in the same cycle.\footnote{The reason is stated right after (\ref{eq:mouse0416}), i.e., it has the highest dimension of $N$ in the combinatorics of $n_i$'s.}
This assignment of $n_i$'s leads to that partitions of $\{k_1,\cdots,k_m\}$ and
$\{n_1,\cdots,n_m\}$ in (\ref{eq:mouse0412}) are Kreweras
complementation map of each other.
The contribution of
$\varpi$ to (\ref{eq:mouse0412}) can be obtained from its $K$-graph as well as the integer assignments of $k_i$'s and $n_i$'s, given by
\begin{eqnarray}\label{eq:cow0417}
&&K^{-1}N^{-m}\mathop{\sum_{b_1,\cdots,b_{|\varpi|}}}_{b_1\neq\cdots\neq b_{|\varpi|}}
\mathop{\sum_{c_1,\cdots,c_{m-|\varpi|+1}}}_{c_1\neq \cdots\neq c_{m-|\varpi|+1}}
\prod_{s_1\in T_1}V(c_{s_1},b_1)\cdots\prod_{s_{|\varpi|}\in
T_{|\varpi|}}V(c_{s_{|\varpi|}},b_{|\varpi|}),
\end{eqnarray}
where $T_i$, $1\leq i\leq |\varpi|$, is the set composed of the
indices of cycles that vertex $i$ resides. For example, for the
noncrossing partition associated with Fig.~\ref{fig:mouse0411}, we
have $T_1=\{1,4,5\}$, $T_2=\{1,2,3\}$ and $T_3=\{4\}$.

We have the following key observation. Suppose that the $i$-th block
of $\varpi$ contains $k$ elements and is given by $B_i=\{B_i(1),\cdots,B_i(k)\}$. Then,
all the edges $\{\overline{B_i(1)},\cdots,\overline{B_i(k)}\}$ in the $K$-graph must be located in
distinct cycles, and these $k$ cycles are
exactly the cycles that vertex $i$ resides. For instance, the
noncrossing partition in Fig.~\ref{fig:mouse0411} has
$B_1=\{1,5,7\}$. It is seen that edges $\overline{1},\overline{5}$
and $\overline{7}$ are in distinct cycles $C_1, C_4$ and $C_5$,
respectively, and these three cycles are the ones that vertex $B_1$
is located. Since the edges partitioning of
$\{\overline{1},\cdots,\overline{m}\}$ is the Kreweras
complementation map of $\varpi$ for $\{1,\cdots,m\}$, the index of
the cycle that edge $\overline{r}$ is located can be identified by
${\cal B}_{K(\varpi)}(r)$ (defined in Definition~\ref{def:3}). Thus,
given $B_i=\{B_i(1),\cdots,B_i(|B_i|)\}$, the set $T_i$ in
(\ref{eq:cow0417}) can be expressed as
\begin{eqnarray}
T_i&=&\{{\cal B}_{K(\varpi)}(B_i(1)),\cdots,{\cal
B}_{K(\varpi)}(B_i(|B_i|)\}\nn\\
&=&\{{\cal B}_{K(\varpi)}(t):t\in B_i\},\quad 1\leq i\leq
|\varpi|.\nn
\end{eqnarray}
It follows that (\ref{eq:cow0417}) can be rewritten as (\ref{eq:tiger0417}) shown
in the following lemma.
\begin{lemma}\label{lemma:1}
Suppose that $\varpi=\{B_1,\cdots,B_{|\varpi|}\}$ is a noncrossing partition of the totally ordered set $\{k_1,\cdots,k_m\}$. Let $k_i$'s in (\ref{eq:mouse0412}) take the same integer in $[1,K]$ if and only if they are partitioned in the same block of $\varpi$. Then, the sum of (\ref{eq:mouse0412}) with the highest order of $N$ is given by
\begin{eqnarray}\label{eq:tiger0417}
&&K^{-1}N^{-m}\mathop{\sum_{b_1,\cdots,b_{|\varpi|}\in[1,K]}}_{b_1\neq\cdots\neq b_{|\varpi|}}
\mathop{\sum_{c_1,\cdots,c_{m-|\varpi|+1}\in[1,N]}}_{c_1\neq\cdots\neq c_{m-|\varpi|+1}}
\prod_{i=1}^{|\varpi|}\prod_{s_i\in\{{\cal B}_{K(\varpi)}(t):t\in
B_i\}}V(c_{s_i},b_i).
\end{eqnarray}
\end{lemma}
\begin{proof}
The proof is straightforward from the preceding discussion.
\end{proof}
We evaluate (\ref{eq:tiger0417}) asymptotically. Rewrite
(\ref{eq:tiger0417}) as
\begin{eqnarray}\label{eq:cow0827}
&&K^{|\varpi|-1}N^{-|\varpi|+1}\times\nn\\
&&\sum_{c_1,\cdots,c_{m-|\varpi|+1}}N^{-m+|\varpi|-1}
\prod_{i=1}^{|\varpi|}\left\{\sum_{b_i}
K^{-1}\prod_{s_i\in\{{\cal B}_{K(\varpi)}(t):t\in
B_i\}}v^{(N)}\left(\dfrac{c_{s_i}-1}{N},\dfrac{b_i-1}{K}\right)\right\},
\end{eqnarray}
where the term inside the bracket $\{\cdot\}$ is identified as the Riemann sum of
$$
\prod_{s_i\in\{{\cal B}_{K(\varpi)}(t):t\in
B_i\}}v^{(N)}\left(\dfrac{c_{s_i}-1}{N},y_i\right)
$$
in $y_i\in[0,1]$ with the partition and tag both being
$\{i/K: 0\leq i\leq K-1\}$. Similarly, the second line of (\ref{eq:cow0827}) is the multi-dimensional Riemann
sum of
$$
\prod_{i=1}^{|\varpi|}\int_0^1 \prod_{s_i\in\{{\cal
B}_{K(\varpi)}(t):t\in B_i\}} v^{(N)}(x_{s_i},y_r)\textrm{d}y_i
$$
on $\{x_1,\cdots,x_{m-|\varpi|+1}\}\in[0,1]^{m-|\varpi|+1}$ with the
partition and tag on each dimension $\{i/N:0\leq i\leq N-1\}$.
Evaluating (\ref{eq:cow0827}) asymptotically, i.e., $K,N\to\infty$ and $K/N\to\beta$, we obtain
\begin{equation}\label{eq:mouse0827}
\beta^{|\varpi|-1}\int_0^1\cdots\int_0^1 \left\{\prod_{i=1}^{|\varpi|}\int_0^1 \prod_{s_i\in \{{\cal
B}_{K(\varpi)}(t):t\in B_i\} }v(x_{s_i},y_i)\textrm{d}
y_i\right\}\textrm{d}x_1\textrm{d}x_2\cdots
\textrm{d}x_{m-|\varpi|+1}.
\end{equation}
To get $\hat{\mu}_m$, we sum up (\ref{eq:mouse0827}) for all
$\varpi\in NC(m)$ and multiply the sum by $\beta$ (due to (\ref{eq:mouse0712})). By Theorem~1.1 of \cite{wachter78} (or Lemma~2.60 of \cite{tulino2004} in brief), we know $\mu_m^{(N)}$ converges a.s. to $\hat{\mu}_m$.

\subsection{Proofs for $\hat{\eta}_{m,k}$ and $\hat{\delta}_{m,k}$}

To show that $\hat{\eta}_{m,k}$ is equal to (\ref{eq:mouse0501}), we set variable $b_1$ in (\ref{eq:tiger0417}) equal to $k$ and multiply (\ref{eq:tiger0417}) by $K$. By writing the resultant equation in the form of Riemann sum and taking the limit $K,N\rightarrow\infty$ and $K/N\to\beta$, we can obtain (\ref{eq:mouse0501}).

Next, we show that $\hat{\delta}_{m,k}$ is equal to (\ref{eq:tiger0827}). The operand of the limit in the expression of $\hat{\delta}_{m,k}$ is given as
\begin{equation}\label{eq:cow0712}
\mathop{\sum_{k_2,\cdots,k_{m+1}\in[1,K]\setminus\{k\}}}_{n_1,\cdots,n_{m+1}\in[1,N]} \textrm{E}\{H_{n_1,k}^* H_{n_1,k_2} H_{n_2,k_2}^* H_{n_2,k_3}\cdots H_{n_{m+1},k_{m+1}}^* H_{n_{m+1},k}\}.
\end{equation}
Since $k_2,\cdots,k_{m+1}$ are unequal to $k$, we should let $n_1=n_{m+1}$; otherwise, the summand is zero. Thus,
(\ref{eq:cow0712}) can be further written as
\begin{equation}
N^{-1}\mathop{\sum_{k_2,\cdots,k_{m+1}\in[1,K]\setminus\{k\}}}_{n_1,\cdots,n_{m}}
V(n_1,k) \textrm{E}\left\{\left(H_{n_1,k_2}
H_{n_2,k_2}^*\right)\cdots \left(H_{n_{m},k_{m+1}}
H_{n_1,k_{m+1}}^*\right)\right\}.\label{eq:dragon0430}
\end{equation}
Let us change the variable $k_{m+1}$ as $k_1$. Then (\ref{eq:dragon0430}) becomes
\begin{equation}\label{eq:mouse1002}
N^{-1}\mathop{\sum_{k_1,\cdots,k_{m}\in[1,K]\setminus\{k\}}}_{n_1,\cdots,n_{m}}
V(n_1,k) \textrm{E}\left\{\left(H_{n_m,k_1}
H_{n_1,k_1}^*\right)\left(H_{n_1,k_2}
H_{n_2,k_2}^*\right)\cdots \left(H_{n_{m-1},k_{m}}
H_{n_m,k_{m}}^*\right)\right\},
\end{equation}
where the term inside expectation is the same as that of (\ref{eq:mouse0412}) except for a conjugate operation. Using a similar reasoning as we have proceeded to obtain (\ref{eq:tiger0417}), we can write (\ref{eq:mouse1002}) as
\begin{equation}
N^{-(m+1)}\mathop{\sum_{b_1,\cdots,b_{|\varpi|}}}_{b_1\neq\cdots\neq b_{|\varpi|}}
\mathop{\sum_{c_1,\cdots,c_{m-|\varpi|+1}}}_{c_1\neq\cdots\neq c_{m-|\varpi|+1}}
V(c_1,k) \prod_{i=1}^{|\varpi|}\prod_{s_i\in\{{\cal B}_{K(\varpi)}(t):t\in
B_i\}}V(c_{s_i},b_i),\label{eq:horse0430}
\end{equation}
with $b_1,\cdots,b_{|\varpi|}$ unequal to $k$. Writing
(\ref{eq:horse0430}) as a Riemann sum, letting
$N\rightarrow\infty$, and summing up for all $\varpi\in NC(m)$, we can obtain (\ref{eq:tiger0827}).

The proof of the a.s. convergence of random sequences $\eta_{m,k}^{(N)}$ and $\delta_{m,k}^{(N)}$ follows from Theorem~1.1 of \cite{wachter78}.

\section{Proof of Theorem~\ref{corollary:1}}\label{appendix:3}

Part 1) of this theorem has been proved in \cite{li04}. Since the proofs of Parts 2) and 3) are very similar, here we only  prove Part 2). The proof of Part 3) simply follows the same line.

Consider $\varpi\in NC(m)$ such that $\varpi=\{B_1,\cdots,B_l\}$ and $K(\varpi)=\{C_1,\cdots,C_{m-l+1}\}$.
When $v(x,y)=g(x)h(y)$ and $v_k(x)=\alpha_k g(x)$, due to the property of noncrossing partition, the contribution of $\varpi$ to (\ref{eq:mouse0501}) can be written as
\begin{equation}\label{eq:mouse0714}
\beta^{l-1}\alpha_k^{|B_1|}\prod_{i=2}^{l}\textrm{E}\{h(Y_i)^{|B_i|}\}
\prod_{j=1}^{m-l+1}\textrm{E}\{g(X_j)^{|C_j|}\}.
\end{equation}
Define a permutation operator $\cal P$ that yields a nonascending order sequence. For all $\varpi=\{B_1,\cdots,B_l\}\in NC(m)$ such that ${\cal P}(|B_1|,\cdots,|B_l|)=(b_1,\cdots,b_l)$, the ratio of $\varpi$'s having $|B_1|=b_n$ is equal to $b_n/(b_1+\cdots+b_l)=b_n/m$.
It is known that the number of $\varpi\in NC(m)$ meeting conditions of
\begin{itemize}
\item $\varpi$ has $l$ blocks with sizes in a nonascending order of $(b_1,\cdots,b_l)$, and

\item $K(\varpi)$ has $m-l+1$ blocks with sizes in a nonascending order of $(c_1,\cdots,c_{m-l+1})$,
\end{itemize}
is $m(m-l)!(l-1)!/f(b_1,\cdots,b_l)/f(c_1,\cdots,c_{m-l+1})$ [\citenum{biane96},\citenum{li04}]. Thus, it is straightforward to see that (\ref{eq:mouse0829}) holds.

\section{Proof of Theorem~\ref{theorem:recursion}}\label{appendix:recursion}

\subsection{Proof for $\hat{\mu}_m$}

For $m\geq 1$ and $1\leq i\leq m$, let $NC^{(i)}(m)$ denote the set
of noncrossing partitions in $NC(m)$ such that the block containing 1
contains $i$ as its largest element. Because of the noncrossing
condition, a partition $\varpi\in NC^{(i)}(m)$ can be decomposed
into $\varpi=\varpi_1 \cup\varpi_2$, where $\varpi_1 \in
NC^{(i)}(i)$, $\varpi_2\in NC(\{i+1,\cdots,m\})$, and $|\varpi|=|\varpi_1|+|\varpi_2|$. It is clear that $NC^{(i)}(i)$ is in bijection with $NC(i-1)$, so we can let
$\varpi_1\in NC(\{1,\cdots,i-1\})$.

Consider a noncrossing partition $\varpi\in NC^{(i)}(m)$, and
$\varpi=\varpi_1 \cup\varpi_2$ is the same as those described above. In its $K$-graph,
there are $|\varpi_1|+|\varpi_2|$ vertices, where vertices $1$ to
$|\varpi_1|$ correspond to blocks of $\varpi_1$, and the
$(|\varpi_1|+1)$-th to the $(|\varpi_1|+|\varpi_2|)$-th vertices
correspond to blocks of $\varpi_2$. Similarly, among the
$m-|\varpi_1|-|\varpi_2|+1$ cycles, cycles $1$ to
$(i-|\varpi_1|)$ correspond to blocks of $K(\varpi_1)$,
and the remaining cycles correspond to blocks of $K(\varpi_2)$.

To prove the result, the following observations are important. Among
the cycles where vertex 1 is located, there is exactly one (the
$(i-|\varpi_1|+1)$-th cycle) corresponding to $K(\varpi_2)$; all other cycles correspond to $K(\varpi_1)$. For any other vertex (other than vertex $1$), either all its located cycles correspond to $K(\varpi_1)$ or all the cycles correspond to $K(\varpi_2)$, but never some to
$K(\varpi_1)$ and some to $K(\varpi_2)$.

Suppose that $\varpi_1=\{D_1,\cdots,D_{|\varpi_1|}\}$ and
$\varpi_2=\{D_{|\varpi_1|+1},\cdots,D_{|\varpi|}\}$. Owing to the above observations, we can write
(\ref{eq:tiger0417}) as
\begin{eqnarray}
&&K^{-1}N^{-m}
\mathop{\sum_{b_{|\varpi_1|+1},\cdots,b_{|\varpi|}}}_{c_{i-|\varpi_1|+1},\cdots,c_{m-|\varpi|+1}}
\prod_{k=|\varpi_1|+1}^{|\varpi|}\prod_{s_k\in\{{\cal
B}_{K(\varpi_2)}(t):t\in D_k\}}V(c_{s_k},b_k)\nn\\
&&\times\mathop{\sum_{b_1,\cdots,b_{|\varpi_1|}}}_{c_1,\cdots,c_{i-|\varpi_1|}}
V(c_{i-|\varpi_1|+1},b_1)
\prod_{l=1}^{|\varpi_1|}\prod_{s_l\in\{{\cal
B}_{K(\varpi_1)}(t):t\in D_l\}}V(c_{s_l},b_l)\nn\\
&=&K^{-1} N^{-m}\sum_{b_1,c_{i-|\varpi_1|+1}}
V(c_{i-|\varpi_1|+1},b_1)
\times
\mathop{\sum_{b_2,\cdots,b_{|\varpi_1|}}}_{c_1,\cdots,c_{i-|\varpi_1|}}
\prod_{l=1}^{|\varpi_1|}\prod_{s_l\in\{{\cal
B}_{K(\varpi_1)}(t):t\in D_l\}}V(c_{s_l},b_l)\nn\\
&&\times
\mathop{\sum_{b_{|\varpi_1|+1},\cdots,b_{|\varpi|}}}_{c_{i-|\varpi_1|+2},\cdots,c_{m-|\varpi|+1}}
\prod_{k=|\varpi_1|+1}^{|\varpi|}\prod_{s_k\in\{{\cal
B}_{K(\varpi_2)}(t):t\in D_k\}}V(c_{s_k},b_k).\label{eq:mouse0801}
\end{eqnarray}
Since $NC(m)=\cup_{i=1}^m NC^{(i)}(m)$ and this is a disjoint union,
we consider all $1\leq i\leq m$ along with all $\varpi_1$ and
$\varpi_2$, which expands (\ref{eq:mouse0801}) to
\begin{eqnarray}
&& N^{-2} \sum_{b_1,c_{i-|\varpi_1|+1}}
V(c_{i-|\varpi_1|+1},b_1)\label{eq:cow0801}\\
&&\times \sum_{i=1}^m N^{-1} N^{-(i-1)+1}\mathop{\sum_{\varpi_1\in
NC(i-1)}}_{\varpi_1=\{D_1,\cdots,D_{|\varpi_1|}\}}
\mathop{\sum_{b_2,\cdots,b_{|\varpi_1|}}}_{c_1,\cdots,c_{i-|\varpi_1|}}
\prod_{l=1}^{|\varpi_1|}\prod_{s_l\in\{{\cal
B}_{K(\varpi_1)}(t):t\in D_l\}}V(c_{s_l},b_l)\nn\\
&&\times K^{-1}N^{-(m-i)+1}\mathop{\sum_{\varpi_2\in
NC(m-i)}}_{\varpi_2=\{D_{|\varpi_1|+1},\cdots,D_{|\varpi|}\}}
\mathop{\sum_{b_{|\varpi_1|+1},\cdots,b_{|\varpi|}}}_{c_{i-|\varpi_1|+2},\cdots,c_{m-|\varpi|+1}}
\prod_{k=|\varpi_1|+1}^{|\varpi|}\prod_{s_k\in\{{\cal
B}_{K(\varpi_2)}(t):t\in D_k\}}V(c_{s_k},b_k).\nn
\end{eqnarray}
Note that there is a slight abuse of notational usage at the first line of (\ref{eq:cow0801}): the summation variable $c_{i-|\varpi_1|+1}$ is indexed by variables $i$ and $\varpi_1$ that appear at the second line. However, as $c_{i-|\varpi_1|+1}$ is a simply dummy variable, this abuse does not affect the result.

Define
$$
U_m(\bar{c},\bar{b}):=N^{-m+1}\mathop{\sum_{\varpi\in NC(m)}}_{\varpi=\{B_1,\cdots,B_{|\varpi|}\}}\mathop{\sum_{b_2,\cdots,b_{|\varpi|}}}_{c_2,\cdots,c_{m-|\varpi|+1}}
\prod_{l=1}^{|\varpi|}\prod_{s_l\in\{{\cal B}_{K(\varpi)}(t):t\in
B_l\}}V(c_{s_l},b_l)\Bigr|_{b_1=\bar{b},c_1=\bar{c}}.
$$
Following the same steps in the proof of Theorem~\ref{theorem:1} for $\hat{\mu}_m$, we have
\begin{equation}\label{eq:tiger0819}
\hat{\mu}_m=\beta\mathop{\lim_{K,N\to\infty}}_{K/N\to\beta}K^{-1}N^{-1}\sum_{\bar{b},\bar{c}}U_m(\bar{c},\bar{b}).
\end{equation}
For each $N$ (with $K/N=\beta$), let $\tilde{\mu}_m^{(N)}$ be a function given by
\begin{equation}\label{eq:rabbit0822}
\tilde{\mu}_m^{(N)}(x,y)=U_m(i,j),\qquad\dfrac{i-1}{N}\leq x<\dfrac{i}{N},\quad
\dfrac{j-1}{K}\leq y<\dfrac{j}{K},
\end{equation}
and the limit of the sequence $\{\tilde{\mu}_m^{(N)}(x,y)\}_{N=1}^\infty$ is $\tilde{\mu}_m(x,y)$. Then, using the Riemann sum expression, (\ref{eq:tiger0819}) can be written as
$$
\hat{\mu}_m=\beta\int_0^1\int_0^1 \tilde{\mu}_m(x,y)\textrm{d}x\textrm{d}y.
$$
Let us represent the expression in (\ref{eq:cow0801}) as $\Omega(\bar{c},\bar{b})$ when the summation at the first line is discarded and $b_1:=\bar{b}$ and $c_{i-|\varpi_1|+1}:=\bar{c}$.
We have
\begin{equation}\label{eq:cow1002}
\beta\mathop{\lim_{K,N\to\infty}}_{K/N\to\beta}\sum_{\bar{b},\bar{c}}\Omega(\bar{c},\bar{b})=\hat{\mu}_m.
\end{equation}
Compare (\ref{eq:tiger0819}) and (\ref{eq:cow1002}) and replace $U_m(\bar{c},\bar{b})$ as $\tilde{\mu}_m^{(N)}((\bar{c}-1)/N,(\bar{b}-1)/K)$, we can see
\begin{equation}\label{eq:dragon0819}
\Omega(\bar{c},\bar{b})=K^{-1}N^{-1} \tilde{\mu}_m^{(N)}\left(\frac{\bar{c}-1}{N},\frac{\bar{b}-1}{K}\right).
\end{equation}
On the other hand, from the expression in (\ref{eq:cow0801}), $\Omega(\bar{c},\bar{b})$ can be also written as
\begin{eqnarray}\label{eq:rabbit0819}
&&N^{-2} v^{(N)}\left(\frac{\bar{c}-1}{N},\frac{\bar{b}-1}{K}\right)\nn\\
&\times&\left[\sum_{i=1}^{m-1}N^{-1}\sum_{c_1} \tilde{\mu}_{i-1}^{(N)}\left(\frac{c_1-1}{N},\frac{\bar{b}-1}{K}\right) \cdot K^{-1}\sum_{b_{|\varpi_1|+1}} \tilde{\mu}^{(N)}_{m-i}\left(\frac{\bar{c}-1}{N},\frac{b_{|\varpi_1|+1}-1}{K}\right)\right.\nn\\
&&\left.+K^{-1}N\cdot N^{-1}\sum_{c_1} \tilde{\mu}_{m-1}^{(N)}\left(\frac{c_1-1}{N},\frac{\bar{b}-1}{K}\right)\right],
\end{eqnarray}
where the sum $\sum_{i=1}^m$ in (\ref{eq:cow0801}) is decomposed as $1\leq i\leq m-1$ and $i=m$ because $\varpi_2$ at the third line of (\ref{eq:cow0801}) is empty when $i=m$.
Equating (\ref{eq:dragon0819}) and (\ref{eq:rabbit0819}) and taking limits on both sides, we obtain
the recursion in (\ref{eq:snake0819}).

\subsection{Proof for $\hat{\eta}_{m,k}$}

We prove this theorem by following the line of the proof for $\hat{\mu}_m$. First, we derive the formula of $\hat{\eta}_{m,k}$. We multiply (\ref{eq:cow0801}) by $K$ and set $b_1:=k$. Define
\begin{eqnarray}
E_{m,k}(\bar{c}):=N^{-m+1}\mathop{\sum_{\varpi\in NC(m)}}_{\varpi=\{B_1,\cdots,B_{|\varpi|}\}}\mathop{\sum_{b_2,\cdots,b_{|\varpi|}}}_{c_2,\cdots,c_{m-|\varpi|+1}}
\prod_{l=1}^{|\varpi|}\prod_{s_l\in\{{\cal B}_{K(\varpi)}(t):t\in
B_l\}}V(c_{s_l},b_l)\Bigr|_{b_1=k,c_1=\bar{c}}.
\end{eqnarray}
We have
\begin{equation}\label{eq:mouse0822}
\hat{\eta}_{m,k}=\mathop{\lim_{K,N\to\infty}}_{K/N\to\beta}N^{-1}\sum_{\bar{c}}E_{m,k}(\bar{c}).
\end{equation}
For each $N$, let $\tilde{\eta}_{m,k}^{(N)}$ be a function given by
$$
\tilde{\eta}_{m,k}^{(N)}(x)=E_{m,k}(i),\qquad\dfrac{i-1}{N}\leq x<\dfrac{i}{N},
$$
and the limit of the sequence $\{\tilde{\eta}_{m,k}^{(N)}(x)\}_{N=1}^\infty$ is $\tilde{\eta}_{m,k}(x)$. Then, (\ref{eq:mouse0822}) can be written as
$$
\hat{\eta}_{m,k}=\int_0^1 \tilde{\eta}_{m,k}(x)\textrm{d}x.
$$
Following the reasoning to obtaining (\ref{eq:dragon0819}) and (\ref{eq:rabbit0819}), we have the equality of
\begin{eqnarray}\label{eq:cow0822}
&&N^{-2} v_k^{(N)}\left(\frac{\bar{c}-1}{N}\right)\left[K\sum_{i=1}^{m-1}N^{-1}\sum_{c_1} \tilde{\eta}_{i-1,k}^{(N)}\left(\frac{c_1-1}{N}\right) \cdot K^{-1}\sum_{b_{|\varpi_1|+1}} \tilde{\mu}^{(N)}_{m-i}\left(\frac{\bar{c}-1}{N},\frac{b_{|\varpi_1|+1}-1}{K}\right)\right.\nn\\
&&\left.+N\cdot N^{-1}\sum_{c_1} \tilde{\eta}_{m-1,k}^{(N)}\left(\frac{c_1-1}{N}\right)\right]=N^{-1} \tilde{\eta}_{m,k}^{(N)}\left(\frac{\bar{c}-1}{N}\right).
\end{eqnarray}
By taking limits on both sides of (\ref{eq:cow0822}), we acquire the recursion in (\ref{eq:tiger0822}).

\subsection{Proof for $\hat{\delta}_{m,k}$}

Suppose that we have $\tilde{\delta}_{m,k}(x)$ and $\tilde{\mu}_{m}(x)$ such that
$$
\hat{\delta}_{m,k}=\int_0^1 \tilde{\delta}_{m,k}(x)\textrm{d}x\qquad\mbox{and}\qquad \hat{\mu}_m=\beta\int_0^1 \tilde{\mu}_m(x)\textrm{d}x,
$$
where $\tilde{\mu}_m(x)$ can be obtained from $\tilde{\mu}_m(x,y)$ by integrating over $y$. Comparing formulas of $\hat{\mu}_m$ and $\hat{\delta}_{m,k}$ given in Theorem~\ref{theorem:1}, we can see
\begin{equation}\label{eq:dragon0822}
\tilde{\delta}_{m,k}(x)=\beta v_k(x)\tilde{\mu}_m(x).
\end{equation}
Integrating both sides of (\ref{eq:dragon0822}) over $x\in[0,1)$, we obtain the relation of $\hat{\delta}_{m,k}$ and $\tilde{\mu}_m(x,y)$ given in (\ref{eq:horse0822}).

\section{Proof of Theorem~\ref{theorem:ergodicAEM}}\label{appendix:proof5}

Here only the proof for $\mu_m$ is given. The proofs for $\eta_{m,k}$ and $\delta_{m,k}$ can be proceeded in very similar ways. We start the proof from (\ref{eq:tiger0417}), and $\mu_m$ can be written as the sum of
\begin{eqnarray}\label{eq:mouse0924}
\beta\mathop{\lim_{K,N\to\infty}}_{K/N\to\beta}K^{-1}N^{-m}\mathop{\sum_{b_1,\cdots,b_{|\varpi|}}}_{b_1\neq\cdots\neq b_{|\varpi|}}
\mathop{\sum_{c_1,\cdots,c_{m-|\varpi|+1}}}_{c_1\neq\cdots\neq c_{m-|\varpi|+1}}
\textrm{E}\left\{\prod_{i=1}^{|\varpi|}\prod_{s_i\in\{{\cal B}_{K(\varpi)}(t):t\in
B_i\}}V(c_{s_i},b_i)\right\}
\end{eqnarray}
over all noncrossing partitions $\varpi=\{B_1,\cdots,B_{|\varpi|}\}\in NC(m)$, where the expectation is over the distribution of $V(i,j)$. Since, for each $j$, $\{V(i,j):1\leq i\leq N\}$ is a realization of random process $Z_j(u,t)$, and $Z_{j}(u,t)$'s are i.i.d. random processes for distinct $j$, the expectation in (\ref{eq:mouse0924}) is equal to
\begin{equation}\label{eq:cow0924}
\prod_{i=1}^{|\varpi|}\textrm{E}\left\{\prod_{s_i\in \{{\cal B}_{K(\varpi)}(t):t\in
B_i\}}Z(u,c_{s_i})\right\}=\prod_{i=1}^{|\varpi|}\textrm{Mom}_{Z}\Biggl(\bigcup_{s_i\in\{{\cal B}_{K(\varpi)}(t):t\in
B_i\}}\{c_{s_i}\}\Biggr),
\end{equation}
where the moments of $Z(u,t)$ are employed to represent those of any $Z_{b_i}(u,t)$.
Plugging (\ref{eq:cow0924}) back to (\ref{eq:mouse0924}) and expressing the result as a Riemann sum, we obtain
\begin{eqnarray*}
&&\mathop{\lim_{K,N\to\infty}}_{K/N\to\beta}K^{|\varpi|}N^{-|\varpi|}\sum_{c_1,\cdots,c_{m-|\varpi|+1}}N^{-m+|\varpi|-1}
\prod_{i=1}^{|\varpi|}\left\{ \sum_{b_i} K^{-1} \textrm{Mom}_{Z}\Biggl(\bigcup_{s_i\in\{{\cal B}_{K(\varpi)}(t):t\in
B_i\}}\{c_{s_i}\}\Biggr)\right\}\\
&=&\beta^{|\varpi|}\int_0^1\cdots\int_0^1 \prod_{i=1}^{|\varpi|} \textrm{Mom}_{z}\Biggl(\bigcup_{s_i\in\{{\cal B}_{K(\varpi)}(t):t\in
B_i\}}\{x_{s_i}\}\Biggr) \textrm{d}x_1\cdots\textrm{d}x_{m-|\varpi|+1},
\end{eqnarray*}
where the last equation is equal to the right-hand-side of (\ref{eq:tiger0924}) when summed over
all noncrossing partitions $\varpi\in NC(m)$.

\section{Proofs of Theorems~\ref{theorem:4} and \ref{theorem:5}}\label{appendix:4}

\subsection{Proof of Theorem~\ref{theorem:4}}

The proofs for (\ref{eq:mouse0313}), (\ref{eq:mouse0902}) and (\ref{eq:cow0902}) are very similar. Here only (\ref{eq:mouse0313}) is proved. We prove it by means of (\ref{eq:tiger0924}). We have
\begin{equation}\label{eq:tiger0925}
\textrm{Mom}_{z|\pmb X}\left(
{\cal X}_{\varpi;i}\right)=\textrm{E}\Biggl\{\prod_{s_i\in \{{\cal
B}_{K(\varpi)}(t):t\in B_i\} }z(u,X_{s_i})\Biggl|\pmb X \Biggr\}.
\end{equation}
We use $M(u)$, $T(u)$ and $W(u)$ to denote the real numbers that correspond to a sample point $u\in{\cal U}$ for random variables $M$, $T$, and $W$, respectively. It is seen that, for any $u\in{\cal U}$, $\prod_{s_i\in \{{\cal
B}_{K(\varpi)}(t):t\in B_i\} }z(u,X_{s_i})$
is nonzero (equal to $M(u)^{|B_i|}$) if and only if either of the
following conditions holds
\begin{enumerate}
\item $W(u)=1$ \hspace{.5cm} and \hspace{.5cm}
$\max\Bigl\{X_l: l\in\bigcup_{t\in B_i}{\cal
B}_{K(\varpi)}(t)\Bigr\}<T(u)<1$,

\item $W(u)=0$ \hspace{.5cm} and \hspace{.5cm}
$0<T(u)<\min \Bigl\{X_l: l\in\bigcup_{t\in B_i}{\cal
B}_{K(\varpi)}(t)\Bigr\}$.
\end{enumerate}
By considering distributions of $T$ and $W$, events 1) and 2) above
have probabilities
$$
\frac{1}{2}\left(1-\max
{\cal X}_{\varpi;i}
\right)\qquad\mbox{and}
\qquad\frac{1}{2}\min{\cal X}_{\varpi;i},
$$
respectively. Thus, the conditional moment in (\ref{eq:tiger0925}) can be
reduced to
\begin{equation}
\textrm{E}\{M^{|B_i|}\}\cdot\frac{1}{2}\left(
1-\max
{\cal X}_{\varpi;i}+\min{\cal X}_{\varpi;i}\right).\label{eq:tiger1114}
\end{equation}
Plugging (\ref{eq:tiger1114}) back onto (\ref{eq:tiger0924}), we
obtain (\ref{eq:mouse0313}).

\subsection{Proof of Theorem~\ref{theorem:5}}

Only the formula of $\mu_m$ is proved. Similarly to the proof of Theorem~\ref{theorem:4}, $\mu_m$ can be evaluated by means of (\ref{eq:tiger0924}). We can see $\prod_{s_i\in \{{\cal B}_{K(\varpi)}(t):t\in B_i\} }z(u,X_{s_i})$
is nonzero (equal to $M(u)^{|B_i|}$) if and only if
$z(u,X_{s_i})\neq 0$ for all $s_i$. This condition has the
probability of $P(u)^{|B_i|}$. Thus, the conditional moment
$\textrm{Mom}_{z|\pmb X}({\cal X}_{\varpi; i})$ in (\ref{eq:tiger0924}) is reduced to
$\textrm{E}\{M^{|B_i|}\}\textrm{E}\{P^{|B_i|}\}$.
Plugging it back onto (\ref{eq:tiger0924}), we obtain
$$
\mu_m=\mathop{\sum_{\varpi\in
NC(m)}}_{\varpi=\{B_1,\cdots,B_{|\varpi|}\}}\beta^{|\varpi|}
\prod_{i=1}^{|\varpi|}\textrm{E}\left\{M^{|B_i|}\right\}\textrm{E}\left\{P^{|B_i|}\right\},
$$
which is equal to (\ref{eq:tiger0626}) by the fact: the number of $l$-block noncrossing partitions having sizes of blocks in a nonascending order of $b_1,b_2,\cdots,b_l$ is equal to  $m(m-1)\cdots(m-l+2)/f(b_1,b_2,\cdots,b_l)$ \cite{kreweras72}.

\bibliography{Hwang}

\begin{thebibliography}{10}

\bibitem{marcenko67}
V.~A. Mar\u{c}enko and L.~A. Pastur,
\newblock ``{The distribution of eigenvalues in certain sets of random
  matrices},''
\newblock {\em MATH USSR SB}, vol. 1, no. 4, pp. 457--483, 1967.

\bibitem{sayeed02}
A.~M. Sayeed,
\newblock ``{Deconstructing multiantenna fading channels},''
\newblock {\em IEEE Trans. on Signal Processing}, vol. 50, no. 10, pp.
  2563--2579, Oct. 2002.

\bibitem{tulino05_1}
A.~M. Tulino, A.~Lozano, and S.~Verd\'{u},
\newblock ``{Impact of antenna correlation on the capicity of multiantenna
  channels},''
\newblock {\em IEEE Trans. Inform. Theory}, vol. 51, no. 7, pp. 2491--2509,
  July 2005.

\bibitem{weichselberger06}
Werner Weichselberger, Markus Herdin, H{\"{u}}seyin {\"{O}}zcelik, and Ernst
  Bonek,
\newblock ``{A stochastic MIMO channel model with joint correlation of both
  link ends},''
\newblock {\em IEEE Trans. Wireless Commun.}, vol. 5, no. 1, pp. 90--100, Jan.
  2006.

\bibitem{girko90}
V.~L. Girko,
\newblock {\em Theory of Random Determinants},
\newblock Kluwer, 1990.

\bibitem{shlyakhtenko96}
D.~Shlyakhtenko,
\newblock ``{Random Gaussian band matrices and freeness with amalgamation},''
\newblock {\em Int. Math. Res. Note}, vol. 20, pp. 1013--1025, 1996.

\bibitem{tse99}
D.~N.~C. Tse and S.~V. Hanly,
\newblock ``{Linear multiuser receivers: effective interference, effective
  bandwidth and user capacity},''
\newblock {\em IEEE Trans. Inform. Theory}, vol. 45, no. 2, pp. 641--657, March
  1999.

\bibitem{kiran00}
Kiran and N.~C. Tse,
\newblock ``{Effective interference and effective bandwidth of linear multiuser
  receivers in asynchronous CDMA systems},''
\newblock {\em IEEE Trans. Inform. Theory}, vol. 46, no. 4, pp. 1426--1447,
  July 2000.

\bibitem{shamai01}
S.~Shamai and S.~Verd\'{u},
\newblock ``{The impact of frequency-flat fading on the spectral efficiency of
  CDMA},''
\newblock {\em IEEE Trans. Inform. Theory}, vol. 47, no. 4, pp. 1302--1327, May
  2001.

\bibitem{li04}
L.~Li, A.~M. Tulino, and S.~Verd\'{u},
\newblock ``{Design of reduced-rank MMSE multiuser detectors using random
  matrix methods},''
\newblock {\em IEEE Trans. Inform. Theory}, vol. 50, pp. 986--1008, June 2004.

\bibitem{tulino05}
A.~M. Tulino, L.~Li, and S.~Verd\'{u},
\newblock ``{Spectral efficiency of multicarrier CDMA},''
\newblock {\em IEEE Trans. Inform. Theory}, vol. 51, no. 2, pp. 479--505, Feb.
  2005.

\bibitem{tulino2004}
A.~M. Tulino and Sergio Verd\'{u},
\newblock {\em Random Matrix Theory and Wireless Communications}, vol. 1, issue
  1,
\newblock Foundations and Trends in Communications and Information Theory, Now
  Publishers Inc., 2004.

\bibitem{silverstein99}
J.~W. Silverstein,
\newblock ``{Comment: Complements and new developments to \textit{Methodologies
  in spectral analysis of large dimensional random matrices, A review}},''
\newblock {\em Statistica Sinica}, vol. 9, no. 3, pp. 611--677, 1999.

\bibitem{carleman22}
T.~Carleman,
\newblock ``{Sur les s\'{e}ries asymptotiques},''
\newblock {\em Comptes Rendus Acad. Sci., \mbox{Paris 174}}, pp. 1527--1530,
  1922.

\bibitem{golub69}
G.~H. Golub and J.~H. Welsh,
\newblock ``{Calculation of Gauss quadrature rules},''
\newblock {\em Math. Comput.}, vol. 23, no. 106, pp. 221--230, Apr. 1969.

\bibitem{moshavi96}
S.~Moshavi, E.~G. Kanterakis, and D.~L. Schilling,
\newblock ``{Multistage linear receivers for DS-CDMA systems},''
\newblock {\em Int. J. of Wireless Inf. Netw.}, vol. 3, no. 1, pp. 1--17, Jan.
  1996.

\bibitem{muller2001}
R.~R. M{\"{u}}ller and S.~Verd\'{u},
\newblock ``{Design and analysis of low-complexity interference mitigation on
  vector channels},''
\newblock {\em IEEE J. Select. Areas Commun.}, vol. 19, no. 8, pp. 1429--1441,
  Aug. 2001.

\bibitem{tulino01}
A.~M. Tulino and S.~Verd\'{u},
\newblock ``{Asymptotic analysis of improved linear receivers for BPSK-CDMA
  subject to fading},''
\newblock vol. 19, no. 8, pp. 1544--1555, Aug. 2001.

\bibitem{li01}
L.~Li, A.~M. Tulino, and S.~Verd\'{u},
\newblock ``{Asymptotic eigenvalue moments for linear multiuser detection},''
\newblock {\em Communications in Information and Systems}, vol. 1, no. 3, pp.
  273--304, Sept. 2001.

\bibitem{cottatellucci02}
L.~Cottatellucci and R.~R. M{\"{u}}ller,
\newblock ``{Asymptotic design and analysis of multistage detectors with
  unequal powers},''
\newblock in {\em Proc. IEEE Information Theory Workshop (ITW'02)}, Oct. 2002.

\bibitem{hachem02}
W.~Hachem,
\newblock ``{Low complexity polynomial receivers for downlink CDMA},''
\newblock in {\em Proc. Asilomar Conf. on Systems, Signals and Computers}, Nov.
  2002.

\bibitem{voiculescu92}
D.~V. Voiculescu, K.~J. Dykema, and A.~Nica,
\newblock {\em Free Random Variables},
\newblock ser. CRM Monograph Series. Providence, R.I., Amer. Math. Soc., 1992.

\bibitem{tse99_1}
D.~N.~C. Tse,
\newblock ``{Multiuser receivers, random matrices and free probability},''
\newblock in {\em Proc. of 37th Ann. Allerton Conf.}, Sept. 1999.

\bibitem{evans00}
J.~Evans and D.~N.~C. Tse,
\newblock ``{Large system performance of linear multiuser receivers in
  multipath fading channels},''
\newblock {\em IEEE Trans. Inform. Theory}, vol. 46, no. 6, pp. 2059--2078,
  Sept. 2000.

\bibitem{biglieri02}
E.~Biglieri, G.~Taricco, and A.~Tulino,
\newblock ``{Performance of space-time codes for a large number of antennas},''
\newblock {\em IEEE Trans. Inform. Theory}, vol. 48, no. 7, pp. 1794--1803,
  July 2002.

\bibitem{muller02}
R.~R. M{\"{u}}ller,
\newblock ``{A random matrix model for communication via antenna arrays},''
\newblock {\em IEEE Trans. Inform. Theory}, vol. 48, no. 9, pp. 2495--2506,
  Sept. 2002.

\bibitem{debbah03}
M.~Debbah, W.~Hachem, P.~Loubaton, and M.~de~Courville,
\newblock ``{MMSE analysis of certain large isometric random precoded
  systems},''
\newblock {\em IEEE Trans. Inform. Theory}, vol. 49, no. 5, pp. 1293--1311, May
  2003.

\bibitem{muller04}
R.~R. M{\"{u}}ller,
\newblock ``{Random matrices, free probability, and the replica method},''
\newblock in {\em Proc. European Signal Processing Conf.}, Vienna, Austria
  2004.

\bibitem{peacock06}
M.~J.~M. Peacock, I.~B. Collings, and M.~L. Honig,
\newblock ``{Asymptotic spectral efficiency of multiuser multisignature CDMA in
  frequency-selective channels},''
\newblock {\em IEEE Trans. Inform. Theory}, vol. 52, no. 3, pp. 1113--1129,
  Mar. 2006.

\bibitem{ryan07}
{{\O}}. Ryan and M.~Debbah,
\newblock ``{Free deconvolution for signal processing applications},''
\newblock Available on line: http://arxiv.org/abs/cs.IT/0701025.

\bibitem{far06}
R.~Rashidi Far, T.~Oraby, W.~Bryc, and R.~Speicher,
\newblock ``{Spectra of large block matrices},''
\newblock Available on line: http://arxiv.org/abs/cs.IT/0610045.

\bibitem{far08}
R.~Rashidi Far, T.~Oraby, W.~Bryc, and R.~Speicher,
\newblock ``{On slow-fading MIMO systems with nonseparable correlation},''
\newblock {\em IEEE Trans. Inform. Theory}, vol. 54, no. 2, pp. 544--553, Feb.
  2008.

\bibitem{honig2001}
M.~L. Honig and W.~Xiao,
\newblock ``{Performance of reduced-rank linear interference suppression for
  DS-CDMA},''
\newblock {\em IEEE Trans. Inform. Theory}, vol. 47, no. 5, pp. 1928--1946,
  July 2001.

\bibitem{cottatellucci05_1}
L.~Cottatellucci and R.~R. M{\"{u}}ller,
\newblock ``{A systematic approach to multistage detectors in multipath fading
  channels},''
\newblock {\em IEEE Trans. Inform. Theory}, vol. 51, no. 9, pp. 3146--3158,
  Sept. 2005.

\bibitem{kreweras72}
G.~Kreweras,
\newblock ``{Sur les partitions noncrois\'{e}es d'un cycle},''
\newblock {\em Discrete Math.}, vol. 1, pp. 333--350, 1972.

\bibitem{speicher06}
A.~Nica and R.~Speicher,
\newblock {\em Lectures on the Combinatorics of Free Probability},
\newblock London Mathematical Society Lecture Note Series: 335, Cambridge
  University Press, 2006.

\bibitem{xiao00}
W.~Xiao and M.~L. Honig,
\newblock ``{Convergence analysis of adaptive full-rank and multi-stage
  reduced-rank interference suppression},''
\newblock in {\em Proc. Conf. Information Sciences and Systems}, March, 2000.

\bibitem{xiao05}
W.~Xiao and M.~L. Honig,
\newblock ``{Large system transient analysis of adaptive least squares
  filtering},''
\newblock {\em IEEE Trans. Inform. Theory}, vol. 51, no. 7, pp. 2447--2474,
  July 2005.

\bibitem{yin83}
Y.~Q. Yin and P.~R. Krishnaiah,
\newblock ``{A limit theorem for the eigenvalues of product of two random
  matrices},''
\newblock {\em J. Multivar. Anal.}, vol. 13, pp. 489--507, 1983.

\bibitem{tulino06}
A.~M. Tulino, A.~Lozano, and Sergio Verd\'{u},
\newblock ``{Capacity-achieving input covariance for single-user multi-antenna
  channels},''
\newblock {\em IEEE Trans. Wireless Commun.}, vol. 5, no. 3, pp. 662--671,
  March 2006.

\bibitem{hwang06}
C.-H. Hwang,
\newblock ``{Asymptotic spectral distribution of crosscorrelation matrix in
  asynchronous CDMA},''
\newblock {\em submitted to IEEE Trans. Inform. Theory}, Available on line:
  http://arxiv.org/abs/cs.IT/0609076.

\bibitem{wachter78}
K.~W. Wachter,
\newblock ``{The strong limits of random matrix spectra for sample matrices of
  independent elements},''
\newblock {\em The Annals of Probability}, vol. 6, no. 1, pp. 1--18, Feb. 1978.

\bibitem{biane96}
P.~Biane,
\newblock ``{Minimal factorizations of cycle and central multiplicative
  functions on the infinite symmetric group},''
\newblock {\em J. Combin. Theory, Ser. A}, vol. 76, no. 2, pp. 197--212, 1996.

\end{thebibliography}
\bibliographystyle{Hwang}

\end{document}